\documentclass[preprint,12pt]{elsarticle}
\usepackage{}

\usepackage{amsmath}
\usepackage{cases}
\usepackage{hyperref}
\usepackage{amssymb}
\usepackage{graphicx}
\usepackage{subfigure}
\usepackage{pict2e}
\usepackage{bm}

\biboptions{numbers,sort&compress}

\newcommand{\upcite}[1]{\textsuperscript{\textsuperscript{\cite{#1}}}}

\begin{document}
\begin{frontmatter}



\title{Clean Numerical Simulation: A New Strategy to Obtain Reliable Solutions of Chaotic Dynamic Systems}


\author{Xiaomling Li$^3$}
\author{Shijun Liao$^{1,2,3,4}$\corref{cor1}}
\ead{sjliao@sjtu.edu.cn} \cortext[cor1]{Corresponding author.}

\address{
$^1$ State Key Laboratory of Ocean Engineering, Shanghai 200240, China;\\
$^2$ Collaborative Innovative Center for Advanced Ship and Deep-Sea Exploration, Shanghai 200240, China;\\
$^3$ School of Naval Architecture, Ocean and Civil Engineering, Shanghai Jiao Tong University, Shanghai 200240, China;\\
$^4$ Ministry-of-Education Key Laboratory of Scientific and Engineering Computing, Shanghai 200240, China
}

\begin{abstract}
It is well known that chaotic dynamic systems (such as three-body system, turbulent flow and so on) have the sensitive dependence on initial conditions (SDIC).   Unfortunately, numerical noises (such as truncation error and round-off error) always exist in practice.  Thus, due to the SDIC, long-term accurate prediction  of  chaotic dynamic systems is practically impossible.  In this paper, a new strategy for chaotic dynamic systems, i.e. the Clean Numerical Simulation (CNS), is briefly  described, together with its applications to a few Hamiltonian chaotic systems.  With negligible numerical noises, the CNS can provide convergent (reliable) chaotic trajectories in a long enough interval of time.  This is very important for Hamiltonian systems such as three-body problem,  and thus should have many applications in various fields. We find that the traditional numerical methods in double precision cannot give not only reliable trajectories but also reliable Fourier power spectra and autocorrelation functions.  In addition, it is found that  even  statistic  properties of chaotic systems   can not  be correctly obtained by means of  traditional numerical algorithms in double precision, as long as these statistics are time-dependent.  Thus, our CNS results strongly suggest that one had better to  be  very  careful on DNS results of statistically unsteady turbulent flows, although DNS results often agree well with experimental data when turbulent flows are in a statistical stationary state.
\end{abstract}


\end{frontmatter}



\section{Introduction}
Poincar\'{e} discovered in mathematics that some dynamic systems have the sensitivity dependence on initial condition (SDIC) \upcite{1}, say, a tiny difference in initial condition might lead to a significant variation of solution after a long enough time.   The SDIC was rediscovered by Lorenz\upcite{2} in 1960s by means of digit computers, who made the SDIC more popular  by giving it a famous name ``butterfly-effect'',  say, a hurricane in North America might be created by the flapping of the wings of a distant butterfly in South America several weeks earlier.  To make matters even worse,  Lorenz$^{[3\textrm{--}4]}$  further found  that such kind of chaotic simulations are sensitive not only to initial conditions but also to numerical algorithms:   different numerical schemes with different time steps might lead to completely different numerical results of chaos.    Lorenz's conclusions were confirmed by many  researchers$^{[5\textrm{--}7]}$.   This is easy to understand, because numerical noises, i.e. truncation error and round-off error, inherently exist at each time-step for {\em all} numerical schemes, which are enlarged exponentially due to the SDIC of chaos\upcite{2}.   These numerical phenomena lead to the intense arguments$^{[8\textrm{--}9]}$ about the reliability of numerical simulations of chaotic systems.   A few researchers even believed that ``all chaotic responses are simply numerical noise and have nothing to do with the solutions of differential equations''\upcite{8}.   Thus, due to the SDIC (or the butterfly-effect),  it is indeed a challenge  to accurately  simulate chaotic solution of nonlinear dynamic systems in a long interval of time$^{[10\textrm{--}11]}$.

It is widely believed that turbulence$^{[12\textrm{--}15]}$ is chaotic.   So,  statistics is commonly used in the study of turbulence, and the direct numerical simulation (DNS)\upcite{16},  which solves the Navier-Stokes equations without averaging or approximation but with all essential scales of motion, has played an important role in turbulence statistics.   However, due to the SDIC, tiny numerical noises, which grow  exponentially in time,  may lead to spurious results.  Yee et al.\upcite{17}  reported   in 1999  that the DNS can produce a spurious solution that is completely different from the  physical solution of their considered case.   In addition,  Wang et al.\upcite{18} demonstrated a spurious evolution of turbulence originated from round-off error in the DNS.  For more examples of spurious numerical simulations, please refer to  Yee et al.$^{[19\textrm{--}20]}$. What's more, analysis of chaotic dynamics were numerically studied in non-linear vibrations of spatial structures such as shells$^{[21\textrm{--}22]}$, plates\upcite{23} and beams$^{[24\textrm{--}26]}$.
Qualitative theory of differential equations is widely used in study of chaotic dynamic systems.
Researchers$^{[23\textrm{--}31]}$ investigated chaotic dynamic systems by means of the Poincar\'{e} maps, Lyanponuv expoent, phase portraits, Fourier and wavelet power spectra, autocorrelation functions, etc. Therefore, it is necessary to verify reliability of these results for chaotic dynamic systems.

Recently,  a new numerical strategy,  namely ``Clean Numerical Simulation" (CNS)$^{[32\textrm{--}42]}$, was  proposed  to  gain  reliable  simulations  of  chaotic dynamic systems in  a  finite  but  long  enough  interval  of  time.   The  CNS  is  based  on an arbitrary Taylor series method (TSM)$^{[43\textrm{--}47]}$ and the multiple precision \upcite{48}  in arbitrary  accuracy,  plus a kind of convergence verification by means of additional simulations with even less numerical noises.   Note that the CNS is a kind of remixing of some known methods, but each of them was used  separately for other purposes.    By means of the CNS, the numerical noises can be  so greatly reduced    to  be much smaller than the  ``true''  solution that the numerical noises are  negligible  in a given interval of time even  for  chaotic  dynamic  systems.   For example, using the traditional Runge-Kutta's method in double precision, one can gain convergent chaotic results of Lorenz equation only in a few dozens of time interval;   however,  using the CNS,  Liao  and  Wang\upcite{35} successfully obtained a convergent, reliable chaotic solution of the Lorenz equation in an interval [0,10000], which is more than 300  times  larger than that given by the traditional Runge-Kutta's method in double precision.
Recently, Li and Liao\upcite{40} demonstrated that 243 more periodic three-body orbits are found by means of the CNS than conventional Runge-Kutta method in double precision and a generalized Kepler's third law for three-body problem was found by Li and Liao \upcite{40} and extended to a generalized Kepler's third law for n-body problem by Sun\upcite{49} with dimensional analysis.
This illustrates the validity of the CNS  for  reliable  simulations  of  chaotic dynamic systems with the SDIC.

In this paper, we further apply the CNS to some chaotic Hamiltonian systems, such as  the H\'{e}non-Heiles system and the famous  three-body problem.  In addition, the influence of numerical noises on statistic results of chaotic dynamic systems  is investigated.  The strategy of the ``Clean Numerical Simulation" (CNS) is briefly described in Section 2.   Applications in chaotic  Hamiltonian systems are given in Section 3.  The influence of numerical noise on chaotic dynamic systems is investigated in Section 4.  The concluding remarks are provided in Section 5.
\section{A brief description of the CNS}

Numerical noises, i.e. truncation and round-off errors, always exist in computer-generated numerical simulations, forever.  In order to gain reliable simulations of chaotic dynamic systems in an arbitrarily long (but finite) interval of time, we must reduce numerical noises to a desired  tiny level.

Consider a nonlinear dynamic system
\begin{equation}
\frac{d\bm{y}}{dt} = \bm{F}(t,\bm{y}), \;\; \bm{y}(0)=\bm{y}_0, \;\;
t\in \mathbb{R}, \;\; \bm{y}_0, \bm{y}, \bm{F} \in \mathbb{R}^n,   \label{geq:CNS}
\end{equation}
where $t$ is the time, $\bm{y}(t)$ is the vector of unknown functions with  $\bm{y}_0$ being its initial value, $\bm{F}$ denotes the vector of nonlinear functions.
Write $t_{n+1}=t_n+h$, where $h$ is the time step.    Then, we have the $M$th-order Taylor expansion
\begin{equation}
\bm{y}(t_{n+1}) \simeq \bm{y}(t_n) + \frac{d\bm{y}(t_n)}{dt}h+\frac{1}{2!}\frac{d^{(2)}\bm{y}(t_n)}{dt^2}h^2+...+\frac{1}{M!}\frac{d^{(M)}\bm{y}(t_n)}{dt^M}h^M,
\label{order:CNS}
\end{equation}
where the Taylor coefficients can be calculated in a recursive way using Eq.(\ref{geq:CNS}).

Obviously, the higher the order $M$ of the Taylor expansion in Eq.(\ref{order:CNS}), the smaller the truncation error, as long as the time-step $h$ is within the radius of convergence.   Besides,  the round-off error can be reduced to any a required level by means of data in multiple precision (MP)\upcite{48} with a large enough number of digits.  Therefore, with enough resources of computation,  one can reduce numerical noises to an arbitrarily tiny level!  In addition,  the convergence (reliability) of a computer-generated result is verified by comparing it to an additional simulation with even smaller numerical noises, a well-known approach widely used in the field of uncertainty quantification.     So, the CNS is based on the remixing of some known methods/technologies.   However, it is interesting that such kind of remixing of the old methods could bring us something new/different, as mentioned below.

\section{Applications of the CNS in some chaotic Hamiltonian Systems}
There are many chaotic Hamiltonian systems in physics, from the microscopic quantum chaos\upcite{50} to the macroscopic solar system\upcite{51}.   Nowadays,   symplectic  integrators (SI)  are widely  applied  to  numerically  solve  Hamiltonian systems$^{[52\textrm{--}54]}$, such as  spin systems\upcite{52}, the solar system\upcite{55}, and so on.   Because the symplectic integrators possess a Hamiltonian as a conserved quantity that often corresponds to the total energy of the system, they have been widely applied to calculate long-term evolution of chaotic Hamiltonian systems with relatively large time steps.

Are long-term orbits of chaotic  Hamiltonian systems obtained by the symplectic integrators indeed reliable?    To check this,  one should be able to gain the convergent chaotic solution in a long enough interval.   Here,  we use the CNS to gain the reliable orbits of the H\'{e}non-Heiles system and the famous  three-body problem,  and compare them with results given by the symplectic integrators in double precision.

\subsection{Symplectic integrator (SI)}

Consider a generic Hamiltonian system
\begin{equation}
	\dot{\bm{p}} = - \frac{\partial H}{\partial \bm{q}},  \;\;\; \dot{\bm{q}} =  \frac{\partial H}{\partial \bm{p}},
\end{equation}
where  $\bm{q}$ denotes the vector of position coordinates,  $\bm{p}$  is  the vector of momentum coordinates,  $H(\bm{p},\bm{q})$ is the Hamiltonian that often corresponds to the total energy of the system.  Assume that the Hamiltonian is separable, say,
\[  H(\bm{p},\bm{q}) = T(\bm{p}) + V(\bm{q}),  \]
where $T(\bm{p})$ is the kinetic energy and $V(\bm{q})$  denotes the potential energy, respectively.    We use here the classical 4th-order explicit symplectic integrator:
\begin{eqnarray}
\bm{q}_i &=& \bm{q}_{i-1}+h \; c_i \; \frac{\partial T(\bm{p}_{i-1})}{\partial \bm{p} },\\
\bm{p}_i &=& \bm{p}_{i-1} - h \; d_i \; \frac{\partial V(\bm{q}_{i})}{\partial \bm{q} },
\end{eqnarray}
for $i=1,2,3,4$, where $h$ denotes the time step and
\begin{eqnarray}
c_1=c_4=\frac{1}{2(2-2^{1/3})}, \;\; c_2=c_3=\frac{1-2^{1/3}}{2(2-2^{1/3})},\nonumber \\
d_1=d_3=\frac{1}{2-2^{1/3}}, \;\; d_2 =-\frac{2^{1/3}}{2-2^{1/3}}, \;\; d_4=0
\end{eqnarray}
are constant coefficients.  For details, please refer to Forest \& Ruth\upcite{56}  and  Yoshida\upcite{57}.

\subsection{The H\'{e}non-Heiles system}

The motion of stars orbiting in a plane about the galactic centre is governed by the so-called  H\'{e}non-Heiles Hamiltonian system of equations\upcite{58}:
\begin{equation}  \label{geq:HH}
\left\{
     \begin{array}{l}
     \ddot{x}(t) = -x(t)-2x(t)y(t),\\
       \\
    \ddot{y}(t) = -y(t)-x^2(t)+y^2(t).
    \end{array}
    \right.
\end{equation}
 Here, the Hamiltonian  is  the total energy, i.e.
 \[ H = T(\dot{x},\dot{y})+V(x,y),  \]
  where
  \[ T(\dot{x},\dot{y})=\frac{1}{2}\left(\dot{x}^2+\dot{y}^2\right)\]
 is  the kinetic energy,
 \[ V(x,y)=\frac{1}{2}\left(x^2+y^2+2x^2y-\frac{2}{3}y^3\right) \]
 is the potential energy, respectively.   As pointed out  by Sprott\upcite{59},  its solution is chaotic for some initial conditions, such as
 \begin{equation}
 x(0)=\frac{14}{25}, \;\;  y(0)=0, \;\;  \dot{x}(0)=0, \;\; \dot{y}(0)=0,  \label{ic:HH}
 \end{equation}
which is considered in this paper.

\begin{figure}
\centering
  \includegraphics[width=8cm]{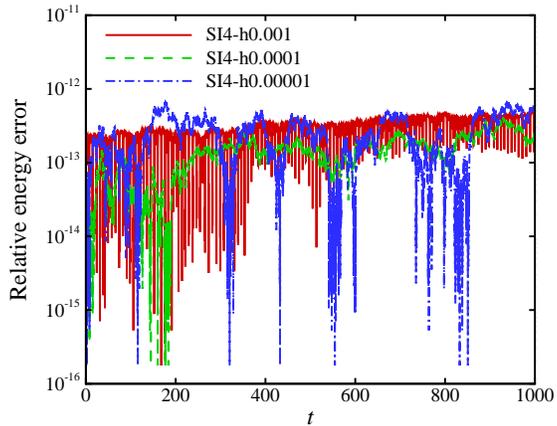}
  \caption{The evolution of relative energy error for the H\'{e}non-Heiles system (see Eqs.(\ref{geq:HH})-(\ref{ic:HH})) given by the 4th-order symplectic integrator using the double precision and the different time steps $h$.  Solid line:  $h=0.001$;  Dashed line: $h = 0.0001$; Dash-dotted line:  $h=0.00001$. }
\label{fig:HH:error}
\end{figure}

\begin{figure}
\centering
  \includegraphics[width=8cm]{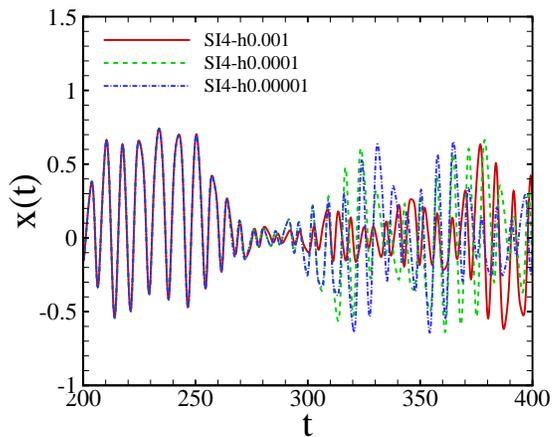}\\
  \caption{The simulations of the $x(t)$ of the H\'{e}non-Heiles system (see Eqs.(\ref{geq:HH})-(\ref{ic:HH})) given by the 4th-order symplectic integrator using the double precision and the different time steps $h$. Solid line:  $h=0.001$;  Dashed line: $h = 0.0001$; Dash-dotted line:  $h=0.00001$.}
\label{fig:HH:x}
\end{figure}

First of all, we simulate the chaotic orbits of the H\'{e}non-Heiles system (Eq.(\ref{geq:HH})) with the initial condition (Eq.(\ref{ic:HH})) in the interval $[0,1000]$  by means of the 4th-order symplectic integrator using the double precision and the several different time steps $h=0.001,0.0001$ and $0.00001$, respectively.  As shown in Fig. \ref{fig:HH:error},  the relative energy errors of all these  simulations given by the 4th-order SI are rather small  in the whole  interval $[0,1000]$,  say,  less than $10^{-12}$.   However,  as shown in Fig.~\ref{fig:HH:x},   their orbits depart from each other quickly:  the orbits given by the SI using $h=0.001$ and $h=0.0001$  separate at about $t= 280$, and the orbits  using  $h=0.0001$ and $h=0.00001$  separate at about $t= 310$, respectively.   Therefore, none of these trajectories  given  by the symplectic integrator are reliable in the interval of $[0,1000]$, even if  the  Hamiltonian, i.e. the total energy of the system, is conserved quite well.

\begin{table*}
\tabcolsep 0pt \caption{The  $x$  and  the deviation from the total energy  at  $t=1000$  by  means  of  the  CNS  using  the different  orders  of  Taylor's  expansion, the data  in the 100-digit multiple precision  and  the  time  step  $h = 1/100$.}\label{table:HH:CNS}
\begin{center}
\def\temptablewidth{1\textwidth}
{\rule{\temptablewidth}{1pt}}
\begin{tabular*}{\temptablewidth}{@{\extracolsep{\fill}}lccc}
Order  & $x(1000)$  & Deviation from the total energy $H$  \\
\hline
20  & -0.049 & $9.1\times 10^{-48}$ \\
25 & -0.049154038397848  & $2.0\times 10^{-60}$ \\
30 & -0.04915403839784844 & $1.8\times 10^{-73}$\\
40 & -0.04915403839784844 & $2.6\times 10^{-96}$\\
\hline
\end{tabular*}
{\rule{\temptablewidth}{1pt}}
\end{center}
\end{table*}

\begin{figure}
\centering
  \includegraphics[width=6cm]{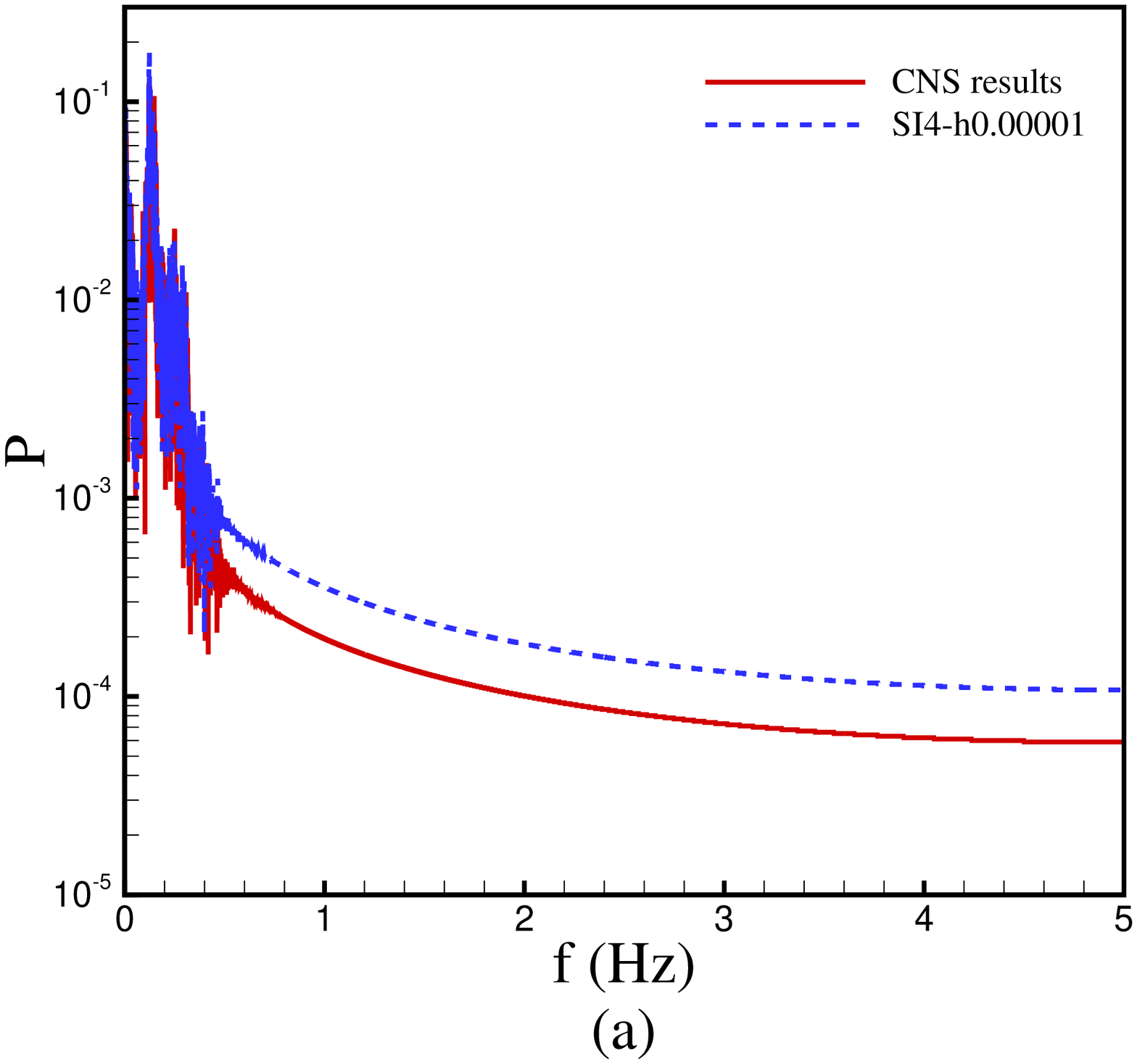}
  \includegraphics[width=6cm]{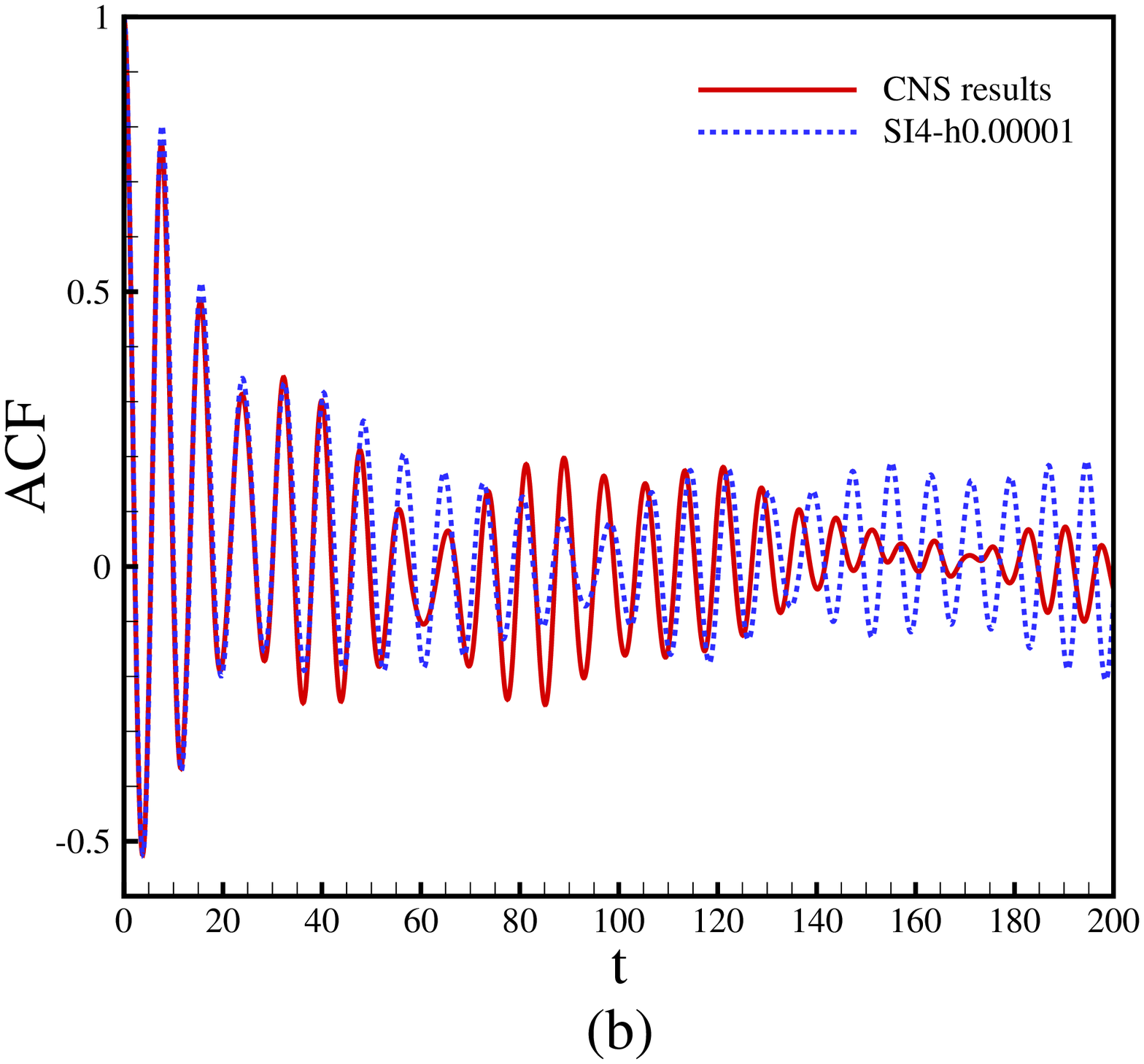}
  \caption{(a) The Fourier power spectra and (b) autocorrelation function (ACF) of the $x(t)$ ($t \in [0,1000]$) of the H\'{e}non-Heiles system given by CNS results and the 4th-order symplectic integrator using the double precision and the time step $h=0.00001$.}
\label{fig:HH:acf}
\end{figure}

Liao\upcite{60} successfully applied the CNS to gain convergent trajectories of the chaotic H\'{e}non-Heiles system (Eq.(\ref{geq:HH})) with the initial condition (Eq.(\ref{ic:HH}))   in the interval $[0,2000]$ by means of the CNS with the 70th-order Taylor expansion and every data in the 140-digit multiple precision using the time step $h =1/10$.   So, following Liao\upcite{60}, we gain a reliable, convergent orbits of Eqs.((\ref{geq:HH})-(\ref{ic:HH})) in the interval [0,1000] by means of the CNS with the 50th-order Taylor expansion and data in the 100-digit multiple precision using the time step $h=1/100$, as listed in Table~\ref{table:HH:CNS}.    Note that the orbits given by the CNS at the 30th-order ($M=30$) of Taylor expansion are convergent in the accuracy of 17 significance digits ,  and the corresponding  deviation  from the total energy  is very small, say,  $1.8\times 10^{-73}$, which is 50 orders of magnitude less than that given by the SI (see Fig.\ref{fig:HH:error}).   Therefore,  unlike the symplectic integrator in double precision,  the CNS can give convergent, reliable orbits of the chaotic H\'{e}non-Heiles system with very small deviation from the total energy (i.e. energy preserving) in a long interval [0,1000].  This  illustrates that the CNS can give more reliable orbits for chaotic  Hamiltonian systems than the SI.
As shown in Fig.~\ref{fig:HH:acf}(a), the Fourier power spectrum of $x(t)$ ($t \in [0,1000]$) of the H\'{e}non-Heiles system given by CNS results is different from the one obtained by the 4th-order symplectic integrator using the double precision and the time step $h=0.00001$ when the frequency $f>0.4$ Hz. This implies that the Fourier power spectrum given by symplectic integrator in double precision is unreliable when the frequency $f>0.4$ Hz. The autocorrelation function of $x(t)$ ($t \in [0,1000]$) of the H\'{e}non-Heiles system obtained by the 4th symplectic integrator has a big difference from the CNS results when $t>50$ as shown in Fig.~\ref{fig:HH:acf}(b). It suggests that the 4th symplectic integrator in double precision cannot give reliable Fourier power spectra and autocorrelation functions.

Why?  This is mainly due to the butterfly-effect of chaos, i.e. the tiny difference at the initial condition enlarges exponentially\upcite{2}.  Here, we should mention that Liao\upcite{60}  applied the CNS to gain a convergent solution of the chaotic  H\'{e}non-Heiles system (Eq.(\ref{geq:HH})) with an initial condition
 \begin{equation}
 x(0)=\frac{14}{25}, \;\;  y(0)=10^{-60}, \;\;  \dot{x}(0)=0, \;\; \dot{y}(0)=0,  \label{ic:HH:2}
 \end{equation}
 which has a tiny difference from Eq.(\ref{ic:HH}), but found that this tiny difference at the initial condition indeed leads to the completely different orbits.  Unfortunately,  the numerical noises (such as the truncation error and round-off error) always exist for the numerical schemes including the symplectic integrators that use the double precision in general with the round-off error at the level $10^{-16}$ that is much less than $10^{-60}$.  So,  it is reasonable that  even the symplectic integrators can {\em not}  give  convergent, reliable long-term trajectories, Fourier power spectra and autocorrelation functions of the chaotic  H\'{e}non-Heiles system in some cases.

\subsection{The three-body problem}

Now,  let us consider another  Hamiltonian system, the  famous  three-body problem,  governed by the Newtonian gravitational law  and the motion equations
\begin{equation}
    \ddot{x}_{k,i}=\sum_{j=1,j\not=i}^3 \frac{Gm_j(x_{k,j}-x_{k,i})}{R_{i,j}^3}, k=1,2,3,  \label{geq:3-body}
\end{equation}
where ${\bf r}_i = (x_{1,i}, x_{2,i},x_{3,i})$ denotes the position of the $i$th body,
$ m_i$ $ (i=1,2,3)$ is the mass of the $i$th body,  and
\begin{equation}
    R_{i,j}=\left [\sum_{k=1}^3\left(x_{k,j}-x_{k,i}\right)^2\right]^{1/2}. \nonumber
\end{equation}
Without loss of generality, let us consider here the case $m_1=m_2=m_3=1$, $G=1$ and  the initial condition
\begin{equation} \label{ic:3-body}
\left\{
\begin{array}{l}
     {\bf r}_1   =  (1/10,0,-1), {\bf r}_2=(0,0,0), {\bf r}_3=(0,0,1), \\
     \\
     \dot{\bf r}_1 = (0,-1,0), \dot{\bf r}_2=(1,1,0), \dot{\bf r}_3=(-1,0,0).
\end{array}
 \right.
\end{equation}

\begin{figure}
\centering
  \includegraphics[width=8cm]{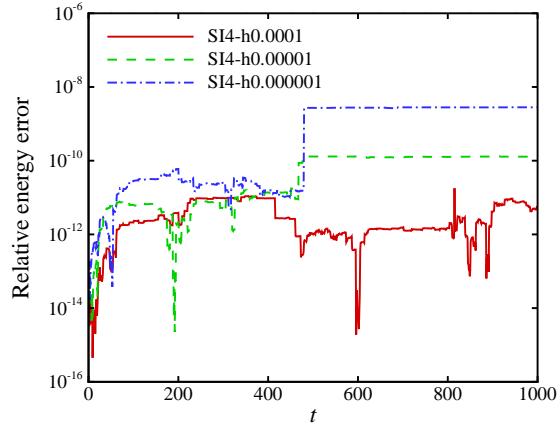}
  \caption{The evolution of the deviation from the total energy of the three-body problem given by the 4th-order symplectic integrator in the double precision using the different time steps $h$.  Solid line: $h=10^{-4}$; Dashed line: $h=10^{-5}$; Dash-dotted line: $h=10^{-6}$.  }
\label{fig:3-body:error}
\end{figure}

\begin{figure}
\centering
  \includegraphics[width=8cm]{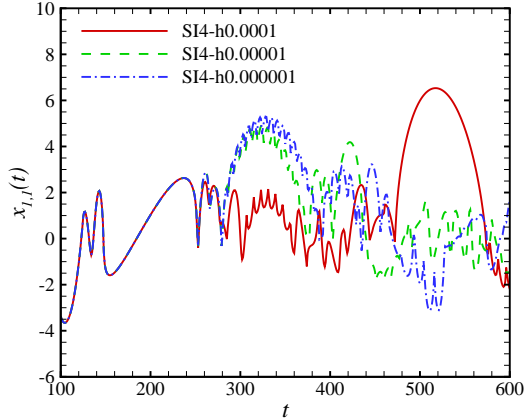}
  \caption{The $x$ position of Body-1 given by the 4th-order symplectic integrator with the double precision using the different time steps $h$.  Solid line: $h=10^{-4}$; Dashed line: $h=10^{-5}$; Dash-dotted line: $h=10^{-6}$. }
\label{fig:3-body:x}
\end{figure}

The 4th-order symplectic integrator in the double precision is used to gain the chaotic orbits of the three-body system (Eqs.(\ref{geq:3-body})-(\ref{ic:3-body})) in the interval [0,1000] by means of the four different time steps $h=10^{-4}, 10^{-5}$ and $10^{-6}$.   Since the three-body problem is a Hamiltonian system,  its total energy must be conserved for a reliable simulation.  As shown in Fig.~\ref{fig:3-body:error},  the  deviation from the total energy of the three-body problem given by the symplectic integrator in double precision is indeed rather small in the {\em whole} interval [0,1000], at a level less than $10^{-8}$.  Unfortunately, this can {\em not} guarantee the reliability of the chaotic trajectory of the three-body system.  A shown in Fig.~\ref{fig:3-body:x},  the $x$ position of the Body-1 given by the time step $h=10^{-4}$ departs at $t\approx 270$ from that given by $h=10^{-5}$, and the $x$ positions given by $h=10^{-5}$ and $h=10^{-6}$ depart from each other at $t\approx 310$, respectively.  Therefore,  the long-term evolutions of the chaotic orbits of the three-body problem  (Eqs.(\ref{geq:3-body})-(\ref{ic:3-body}))  given by the 4th-order symplectic integrator in double precision are {\em not} reliable in the interval [0,1000].

\begin{table*}
\tabcolsep 0pt \caption{The $x$ position of the Body-1 of the three-body system (Eqs.(\ref{geq:3-body})-(\ref{ic:3-body}))  at $t=1000$ given by the CNS with the different orders ($M$) of Taylor expansion, the data in 300-digit multiple precision and the time step $h=10^{-3}$.}\label{table:3-body}
\begin{center}
\def\temptablewidth{1\textwidth}
{\rule{\temptablewidth}{1pt}}
\begin{tabular*}{\temptablewidth}{@{\extracolsep{\fill}}lccc}
Order  & $x_{11}(1000)$  & Deviation from the total energy \\
\hline
40 & -16.8 & $1.63\times 10^{-17}$\\
50 & -16.82869 & $7.88\times 10^{-22}$\\
60 & -16.8286925389 & $2.05\times 10^{-26}$\\
70 & -16.82869253894194 & $1.12 \times 10^{-31}$\\
80 & -16.82869253894194 & $1.90 \times 10^{-35}$\\
\hline
\end{tabular*}
{\rule{\temptablewidth}{1pt}}
\end{center}
\end{table*}

\begin{figure}[ht]
\centering

  \includegraphics[width=6cm]{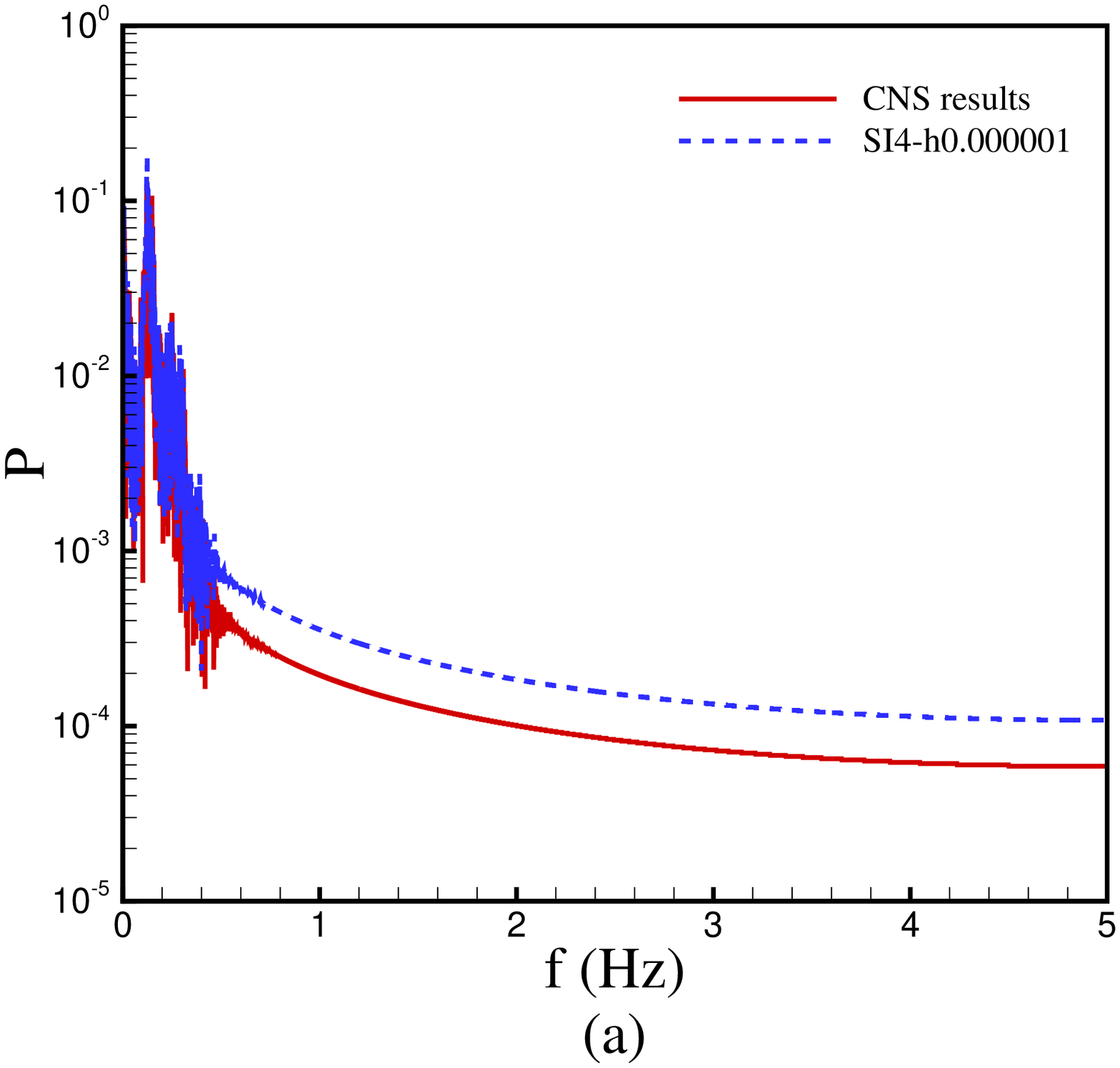}
  \includegraphics[width=6cm]{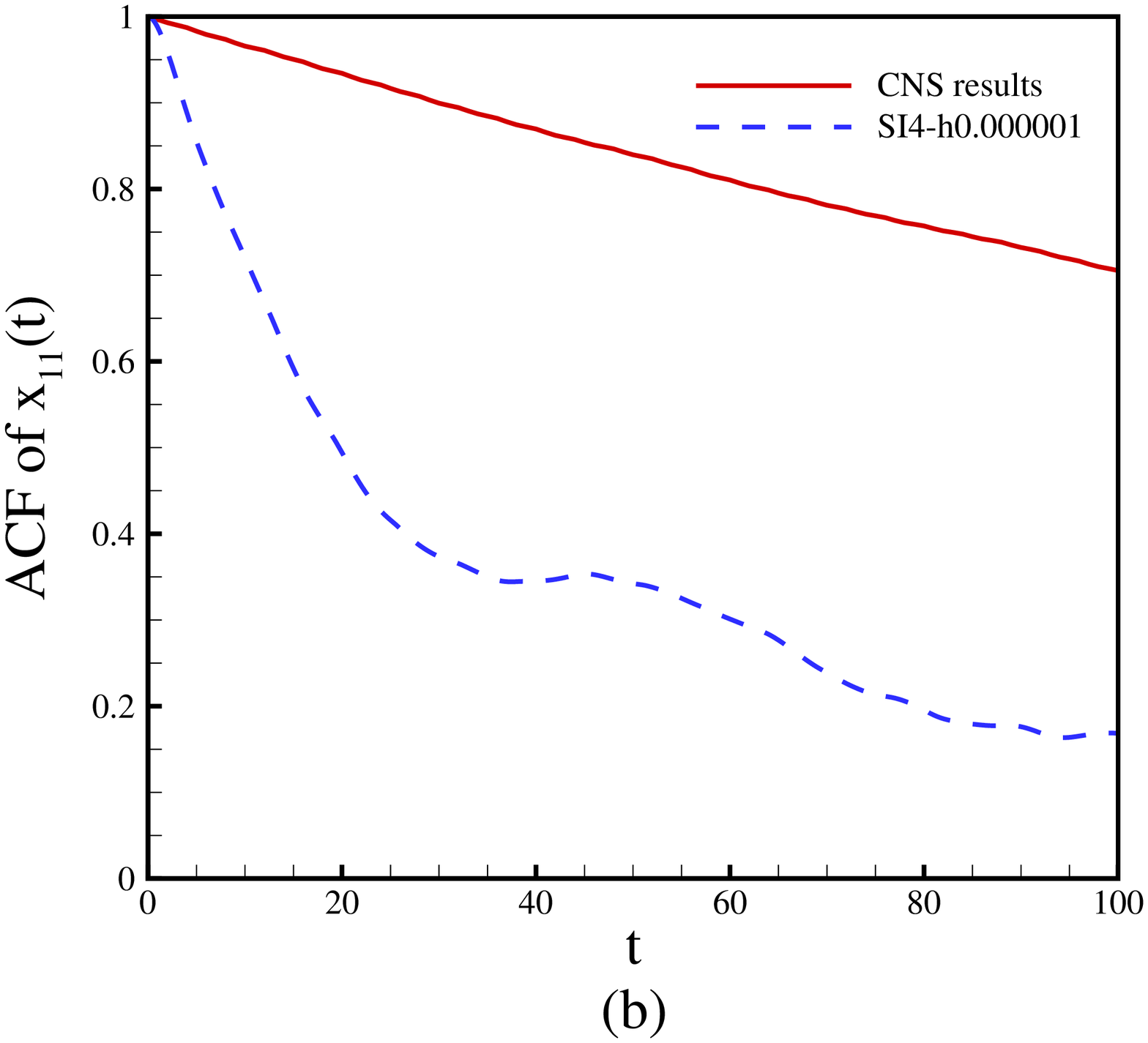}
  \includegraphics[width=6cm]{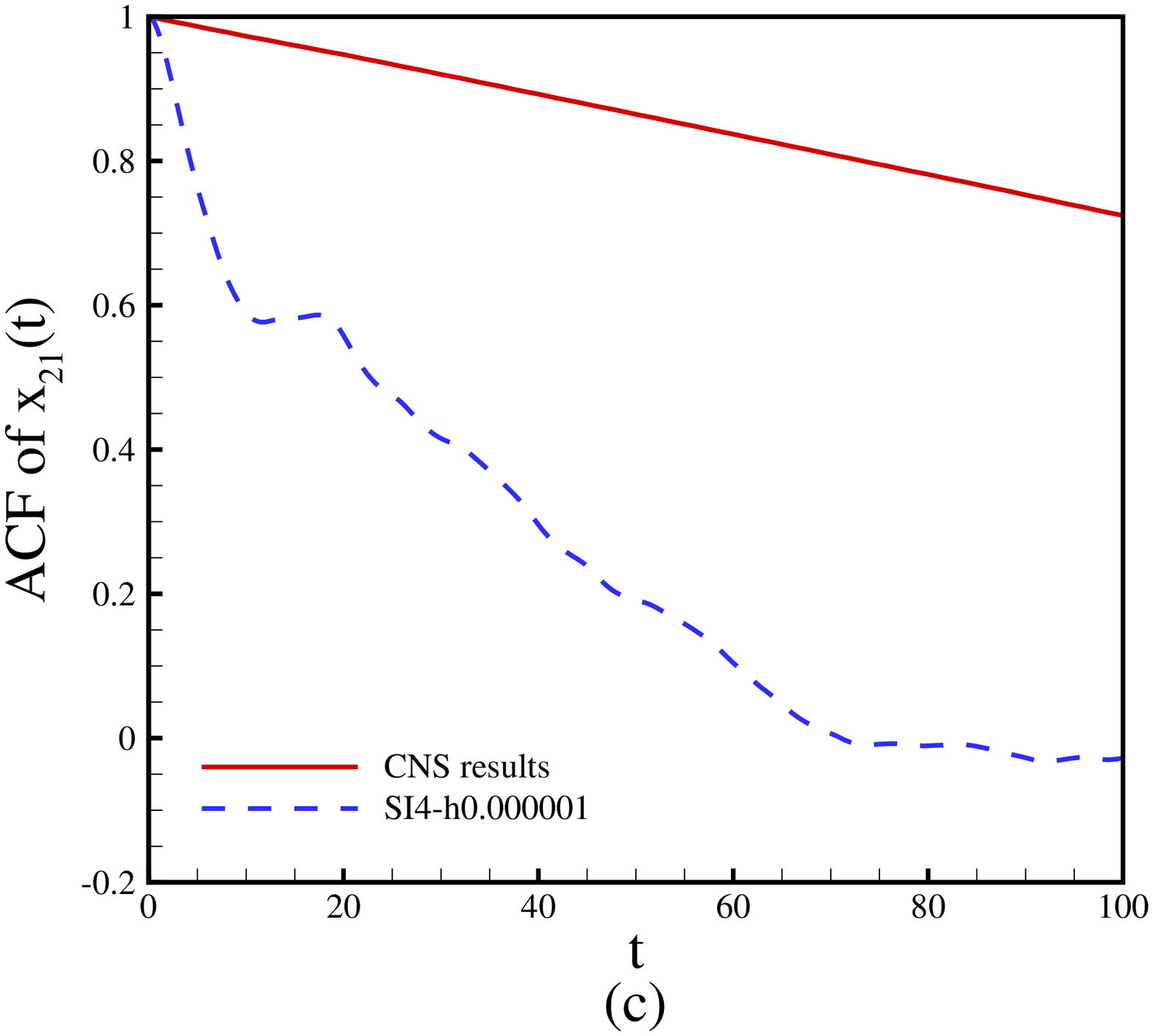}
  \includegraphics[width=6cm]{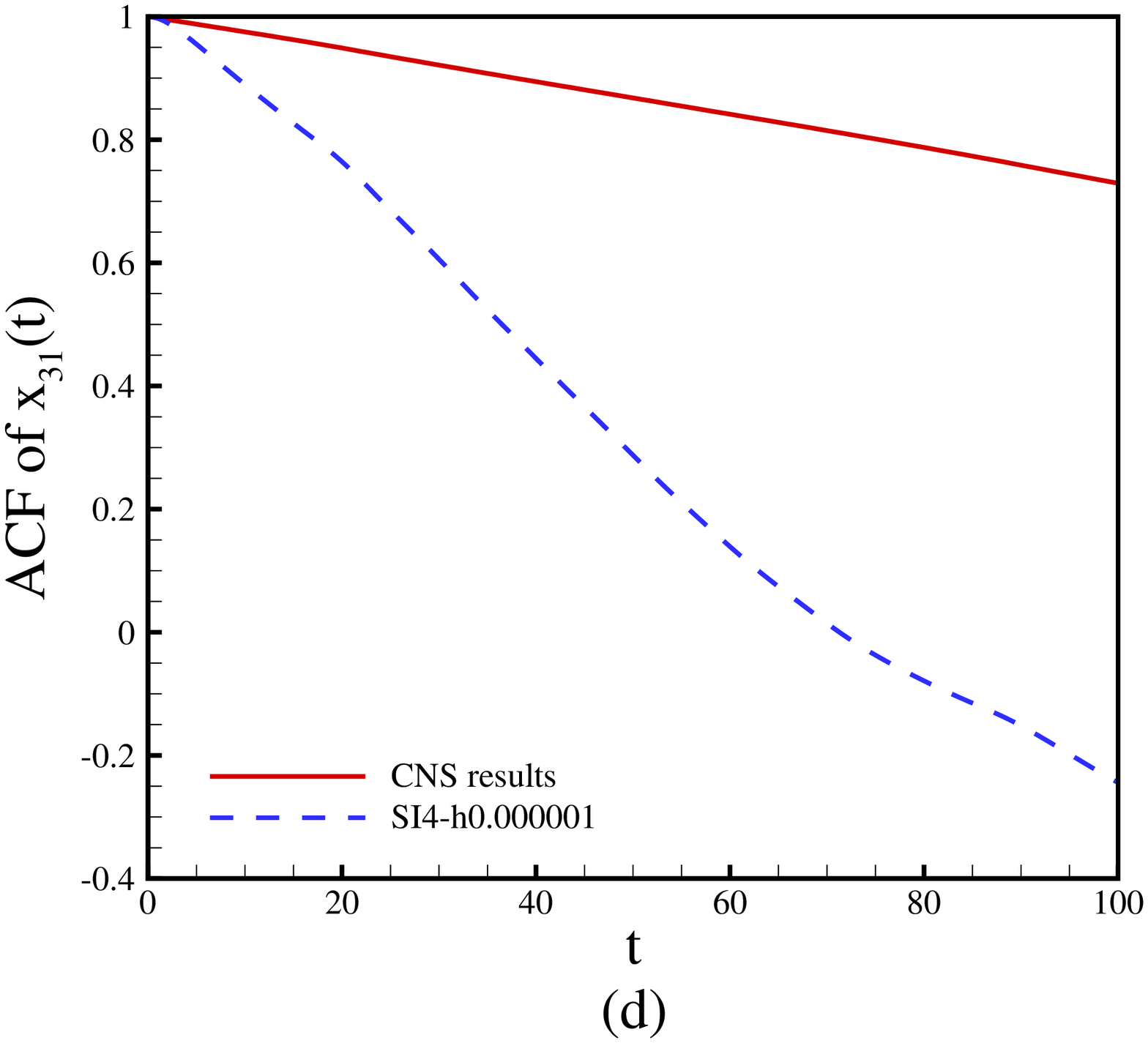}
  \caption{(a) The Fourier power spectra of $x_{11}(t)$ and (b-d) the autocorrelation function (ACF) of the $x_{11}(t)$, $x_{21}(t)$ and $x_{31}(t)$ of Body-1 of the three-body  system given by CNS results and the 4th-order symplectic integrator using the double precision and the time step $h=0.000001$.}
\label{fig:3b:acf}
\end{figure}

By means of the CNS with the 80th-order Taylor expansion, data in 300-digit multiple precision and the time step $h=10^{-2}$,   Liao\upcite{34} successfully gained the reliable orbits of a similar chaotic three-body problem in the interval [0,1000].    Similarly,  following Liao\upcite{34},  we obtained a convergent long-term evolution of the three-body system  (Eqs.(\ref{geq:3-body})-(\ref{ic:3-body})) in the interval [0,1000]  by means of the CNS.     Indeed, by means of the CNS using the up-to 80th-order Taylor expansion and data in the 300-digit multiple precision with the time step $h=10^{-3}$, we gain convergent orbits of the three-body system in the time interval $[0,1000]$, as shown in Table~\ref{table:3-body}.    Note that the orbits at $t=1000$ given by the CNS at the 40th to 60th-order of Taylor expansion are convergent in the accuracy of 3, 7 and 12 significance digits, respectively.  Especially, the orbits given by the CNS at the 70th and 80th-order Taylor expansion are convergent in the accuracy of 16 significance digits, with very small   deviations from the total energy at the level  $10^{-31}$ and $10^{-35}$, respectively.      Thus, unlike the symplectic integrator,  the CNS can give a reliable long-term evolution of the chaotic orbits of the three-body system, together with a rather small deviation from the total energy.   It should be emphasized that, for the three-body problem,  it is  very  important  to  give  an  accurate   prediction of orbits.
As shown in Fig.~\ref{fig:3b:acf}(a), the Fourier power spectrum of the $x_{11}(t)$ ($t\in[0,1000]$) of Body-1 of the three-body  system given by CNS results has large difference from the power spectrum obtained by the 4th-order symplectic integrator using the double precision and the time step $h=0.000001$ when the frequency $f>0.4$ Hz.  This implies that the Fourier power spectrum given by symplectic integrator in double precision is unreliable when the frequency $f>0.4$ Hz. What's worse, the autocorrelation function of the $x_{11}(t)$, $x_{21}(t)$ and $x_{31}(t)$  ($t\in[0,1000]$) of Body-1 of the three-body  system obtained by the 4th symplectic integrator is totally different from the CNS results as shown in Fig.~\ref{fig:3b:acf}~(b), (c) and (d). It also suggests that the 4th symplectic integrator in double precision cannot give reliable Fourier power spectra and autocorrelation functions.

These two examples illustrate that, for some chaotic Hamiltonian systems,  even the symplectic integrators in double precision can {\em not} give convergent (reliable)  numerical simulations of {\em trajectories, Fourier power spectra and autocorrelation functions} in a long enough interval of time. In fact,  to the best of our knowledge, {\em all}  traditional numerical algorithms in {\em double precision} can not give convergent (reliable) long-term prediction of trajectories for chaotic dynamic systems.   But,  the CNS can do it.   So, although the CNS is a remixing of some well-known methods/technologies,  it can indeed bring us something completely new/different!

\section{Influence of numerical noises on statistics of chaotic systems}

It is widely believed that,  although traditional algorithms in double precision can not  give correct trajectories of chaotic systems,    they  however  {\em might}  correctly give statistical properties.  Is this indeed true?   In this section, we illustrate the influence of numerical noise on statistic of chaotic dynamic systems via an example.

It is well known that the famous Lorenz equation\upcite{2} is a quite simplified model of the Rayleigh-B\'{e}nard (RB) flow of viscous fluid.  From the  exact  Navier-Stokes equations for the two-dimensional Rayleigh-B\'{e}nard flow
\begin{eqnarray}
    \nabla^2 \frac{\partial{\psi}}{\partial t} + \frac{\partial{(\psi, \nabla^2\psi)}}{\partial(x,z)} - \sigma \frac{\partial\theta}{\partial x} -
    \sigma \nabla^4\psi = 0 \\
    \frac{\partial\theta}{\partial t} + \frac{\partial(\psi,\theta)}{\partial(x,z)} - R\frac{\partial\psi}{\partial x} - \nabla^2\theta = 0
\end{eqnarray}
where $\psi$ denotes the stream function, $\theta$ the temperature  departure  from a linear variation background,  $t$ the time, $(x,z)$ the horizontal and vertical coordinates, $\sigma$ the Prandtl number, $R$ the Rayleigh number, respectively.  Saltzman\upcite{61}  deduced  a  family of highly  truncated  dynamic systems with different degrees of freedom:  the famous Lorenz equation\upcite{2} is only the simplest  one  among  them.    For example,   in the case of the Prandtl number $\sigma=10$, the highly truncated dynamic system with three degrees of freedom  (three ODEs) reads
\begin{equation}\label{3eqns}
\left\{
\begin{split}
  \dot A &= -148.046A - 1.500D, \\
  \dot D &= -13.958AG - 1460.631\lambda A - 14.805D, \\
  \dot G &= 27.916AD - 39.479G,
\end{split}
\right.
\end{equation}
where $\lambda=R/R_c$ is dimensionless Rayleigh number, $R_c$ is the critical Rayleigh number,  $A$ and $D$ represent the cellular streamline and thermal fields for the Rayleigh critical mode, and $G$ denotes the departure of the vertical variation, respectively.  For details, please refer to  Saltzman\upcite{61} .

Similarly, Saltzman\upcite{61}  gave the highly truncated dynamic system with five degrees of freedom  (five ODEs):
\begin{equation}\label{5eqns}
\left\{
\begin{split}
  \dot A &= 23.521BC - 1.500D - 148.046A, \\
  \dot B &= -22.030AC - 186.429B, \\
  \dot C &= 1.561AB - 400.276C, \\
  \dot D &= -13.958AG - 1460.631\lambda A - 14.805D, \\
  \dot G &= 27.916AD - 39.479G,
\end{split}
\right.
\end{equation}
and that with the seven degrees of freedom (seven ODEs):
\begin{equation}\label{7eqns}
\left\{
\begin{split}
    \dot A &= 23.521BC - 1.500D - 148.046A, \\
    \dot B &= -22.030AC - 1.589E - 186.429B, \\
    \dot C &= 1.561AB - 0.185F - 400.276C, \\
    \dot D &= -16.284CE - 16.284BF - 13.958AG \\
           &\qquad\qquad\qquad   - 1460.631\lambda A -14.805D, \\
    \dot E &= 16.284CD - 16.284AF - 18.610BG \\
           &\qquad\qquad\qquad   -1947.508\lambda B -18.643E, \\
    \dot F &= 16.284AE + 16.284BD - 486.877\lambda C\\
             & \qquad\qquad\qquad  - 40.028F, \\
    \dot G &= 27.916AD + 37.220BE - 39.479G,
\end{split}
\right.
\end{equation}
respectively.   All of them are deterministic equations with chaotic solutions, and are simplified models for the two-dimensional Rayleigh-B\'{e}nard flow.   It should be emphasized  here that, for any a given initial condition, we can gain reliable, convergent numerical results of chaotic solutions of these models in a finite but  long enough interval of time by means of the CNS.

In physics, the two-dimensional Rayleigh-B\'{e}nard flow with  a large  enough Rayleigh number $R$ is an evolutionary process from an initial equilibrium state  to  turbulence after a long enough time, mainly because the flow is unstable and besides the micro-level physical uncertainty (such as thermal fluctuation) always exists.   Such kind of initial micro-level physical uncertainty due to thermal fluctuation can be expressed by Gaussian random data, as illustrated by Wang et al.\upcite{13}.   Mathematically, due to the SDIC,  the chaotic solutions of these simplified models should be dependent upon the random initial conditions.   This is true in physics,  since all experimental measurements of Rayleigh-B\'{e}nard  turbulent flows are different.   However, due to the butterfly-effect, one can not obtain convergent (reliable)  trajectories of these chaotic systems in a long enough interval of time by means of the traditional numerical algorithms in double precision.   Therefore,  the above-mentioned  equations provide us a few simplified models to investigate the influence of numerical errors on the statistics  of  such kind of chaotic dynamic systems.

Without loss of generality, let us first  consider  the  deterministic three ODEs (Eq.(\ref{3eqns})) with the random initial conditions (physically related to the micro-level thermal fluctuation)  in  a normal  distribution  in case of $\lambda = 28$,  corresponding to a turbulent flow.   Considering the thermal fluctuation,  we study here such kind of random  initial  condition in normal distribution  with the mean
\begin{displaymath}
   \langle A(0) \rangle =1, \quad  \langle D(0) \rangle = 10^{-3},\quad   \langle G(0) \rangle = 10^{-3}
\end{displaymath}
and the standard deviation
\begin{equation}
\sigma_{0}  = \sqrt{\langle  A^2(0)\rangle}=\sqrt{\langle D^2(0)\rangle}=\sqrt{\langle G^2(0)\rangle} =10^{-30}.   \nonumber
\end{equation}
Here, the standard deviation  is related to the micro-level thermal fluctuation, which is much smaller than numerical noises of traditional numerical algorithms in double precision.   However,  by means of the CNS with numerical noises much smaller than thermal fluctuation, we can gain convergent (reliable) chaotic propagations (trajectories) of the micro-level physical uncertainty of these systems for any a given initial condition.    Let
\begin{eqnarray}
\langle A(t)\rangle &=&\frac{1}{N}\sum_{i=1}^{N}A_i(t),\\
\sigma_A(t) &=&\sqrt{\frac{1}{N-1}\sum_{i=1}^N[A_i(t)-\langle A(t)\rangle ]^2}
\end{eqnarray}
denote the sample mean and unbiased estimate of standard deviation of $A(t)$ of these reliable simulations of trajectories,  respectively, where $N$ is the number of samples.

\begin{figure}[h]
  \centering
  \includegraphics[scale=0.3]{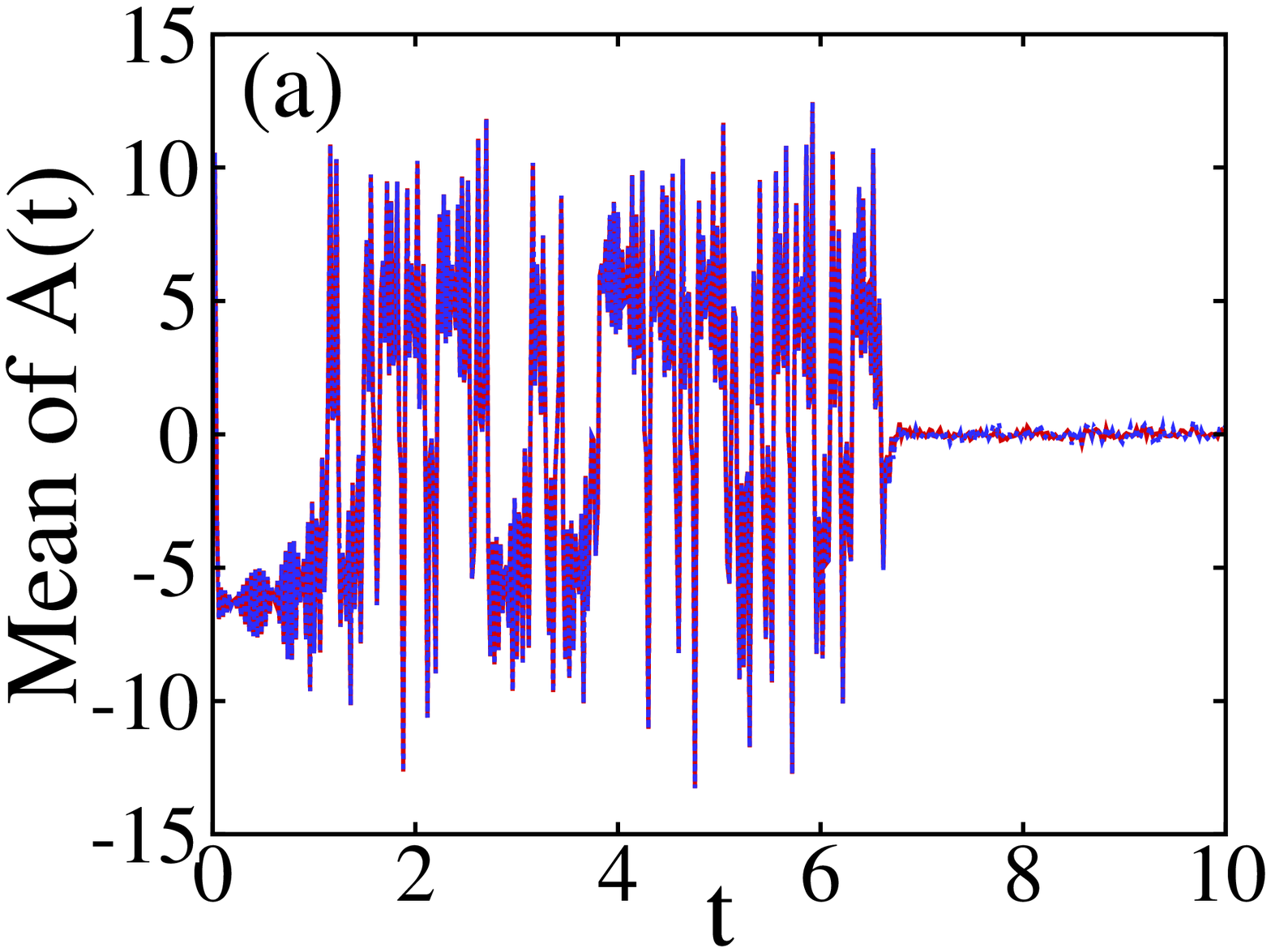}
  \includegraphics[scale=0.3]{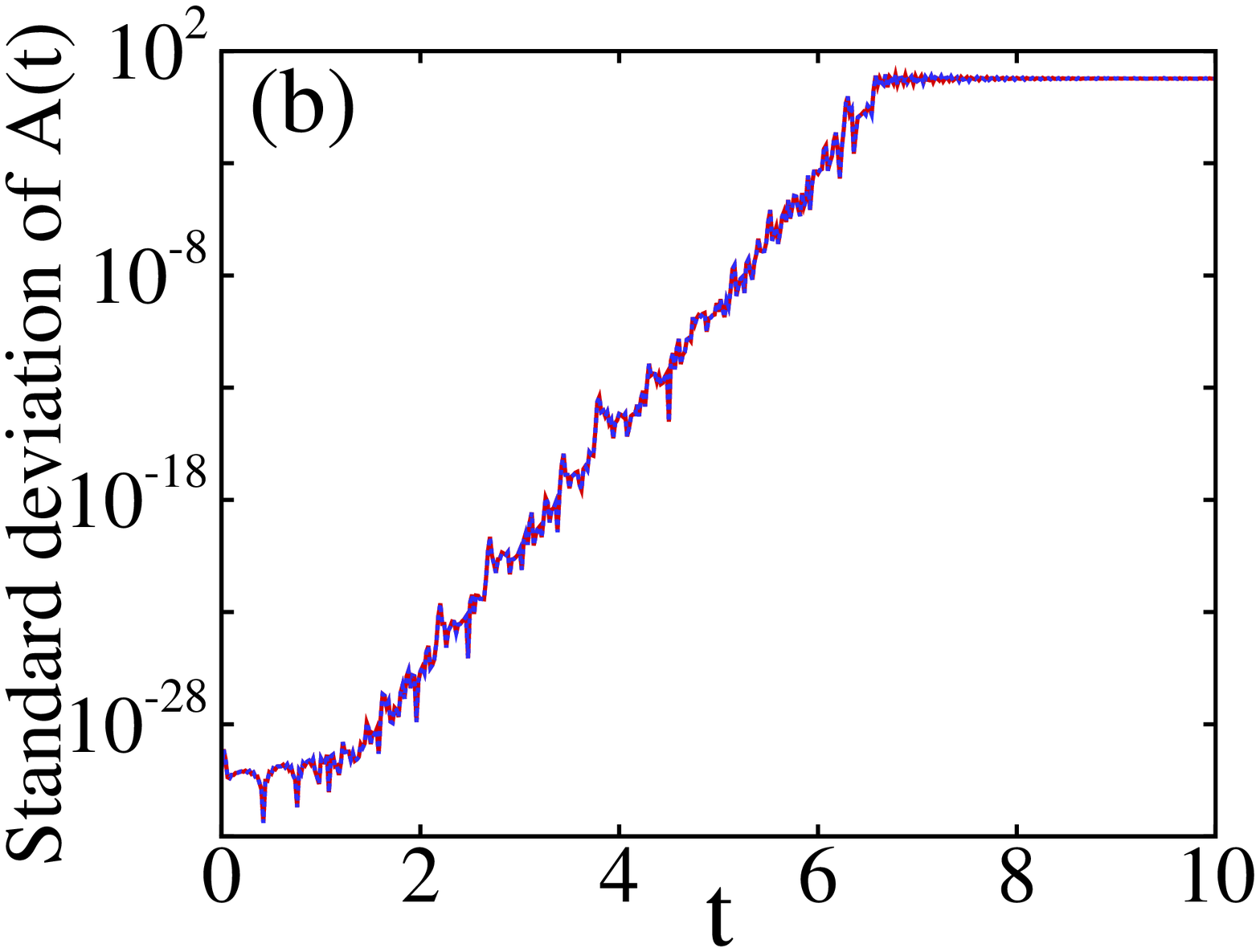}
   \caption{(a) The mean and (b) standard deviation of $A(t)$ of the three ODEs' chaotic system (Eq.(\ref{3eqns}))  using different numbers of samples gained by means of the CNS.  Solid line: 2000 samples; Dashed line: 1000 samples.}
  \label{fig:3eqns-M80K300-2k-1k}
\end{figure}

Two thousand samples of reliable numerical simulations of the three ODEs' system (Eq.(\ref{3eqns})) are obtained in the time interval $[0,10]$ by means of  the CNS using the 80th-order Taylor series ($M=80$),  the 90 decimal-digit precision ($K=90$)  for every data, and the time-step $h =10^{-3}$.   It is found that the numerical errors can be  decreased to be much smaller than the micro-level physical uncertainty in the time interval [0,10] under consideration.    These numerical simulations are so accurate that we can consider them as the ``true'' solutions of the chaotic dynamic system (Eq.(\ref{3eqns})), which can be used to investigate the influence of numerical noises on statistic computations of chaotic systems.    In this way,  we successfully distinguish/separate the ``true''  (convergent)  chaotic solutions that have physical meanings from numerical noises that have not physical meanings!   Note that, for the chaotic dynamic system (Eq.(\ref{3eqns})),  the mean and the standard deviation of $A(t)$ using 1000 samples  are almost the same as those using 2000 samples, as shown in Fig. \ref{fig:3eqns-M80K300-2k-1k}.  Thus, it is enough for us to use 2000 samples in this paper.

\begin{figure}[h]
  \centering
  \includegraphics[scale=0.3]{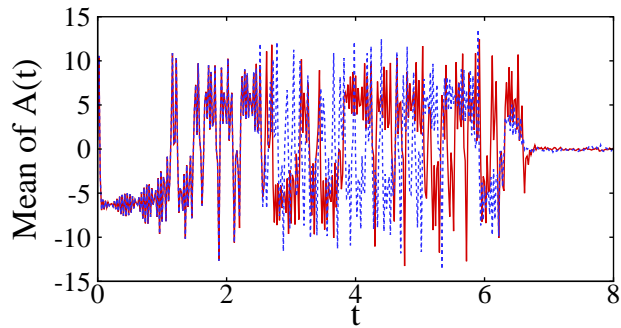}
  \caption{The influence of the truncation error on the mean of $A(t)$ of the three ODEs' chaotic dynamic system (see Eq.(\ref{3eqns})) using the time-step $h=10^{-3}$  and  the 90 decimal-digit precision for every data (i.e. with the negligible round-off error).   Solid line in red:  the reliable  mean of $A(t)$  given by the CNS using $M=80$ (i.e. with the negligible truncation error);  dashed line in blue:  the mean of $A(t)$ given by $M=10$ (i.e. with the considerable truncation error).}
  \label{fig:3eqns-M80K300-M10K300}
\end{figure}

Obviously, the larger the order $M$ of the Taylor series in the frame of the CNS, the smaller the truncation error.   For example,  the traditional 4th-order Runge-Kutta's  method corresponds to the CNS using the 4th-order Taylor series expansion.  Thus, to investigate the influence of the truncation error,  we use here the 10th-order Taylor series, i.e. $M=10$,  but remain the 90 decimal-digit multiple precision for every data so as to make round-off error negligible.       In this way,  the round-off error is negligible in the considered interval of time $t\in[0,10]$ so that the influence of the truncation error can be investigated independently.  Note that our reliable CNS results are gained by means of $M=80$, i.e. the 80th-order Taylor series, whose truncation errors are negligible in the considered interval of time $t\in[0,10]$.  However,  when $M=10$,  the  truncation  error is not  negligible in [0,10].

\begin{figure}
  \centering
  \includegraphics[scale=0.8]{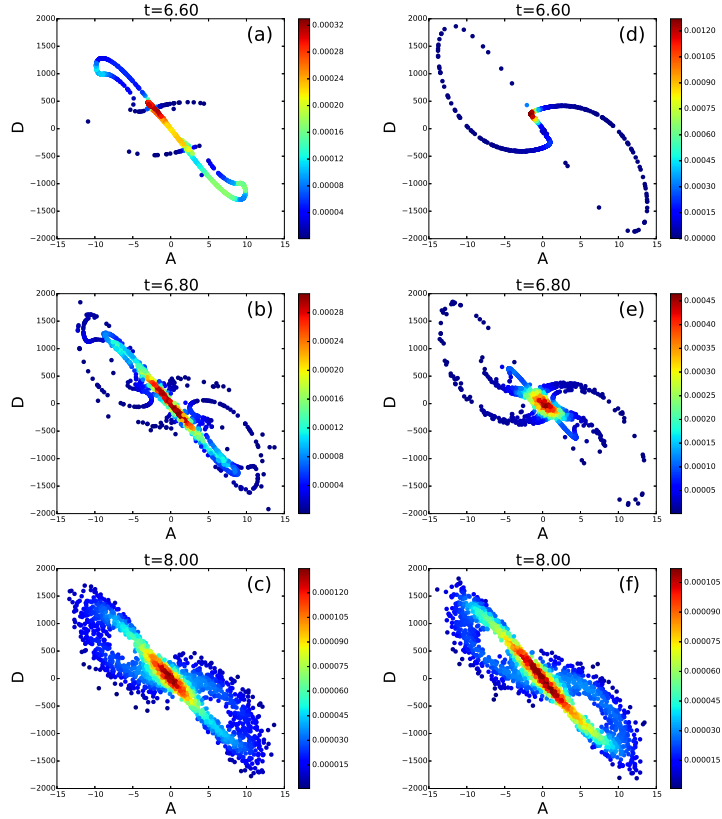}
  \caption{The scatter diagrams of the probability distribution function (PDF) in the $A - D$ plane of the three ODEs' system (see Eq.(\ref{3eqns})) at different times by means of the time-step $h =10^{-3}$,  data in the 90 decimal-digit precision  and  the Taylor series in the different orders $M$.  (a)-(c): unreliable results given by $M=10$; (d)-(f): the reliable  results given by the CNS using $M=80$. (color online)}
  \label{fig:3eqns-GaussianKDE}
\end{figure}

\begin{figure}
  \centering
  \includegraphics[scale=0.3]{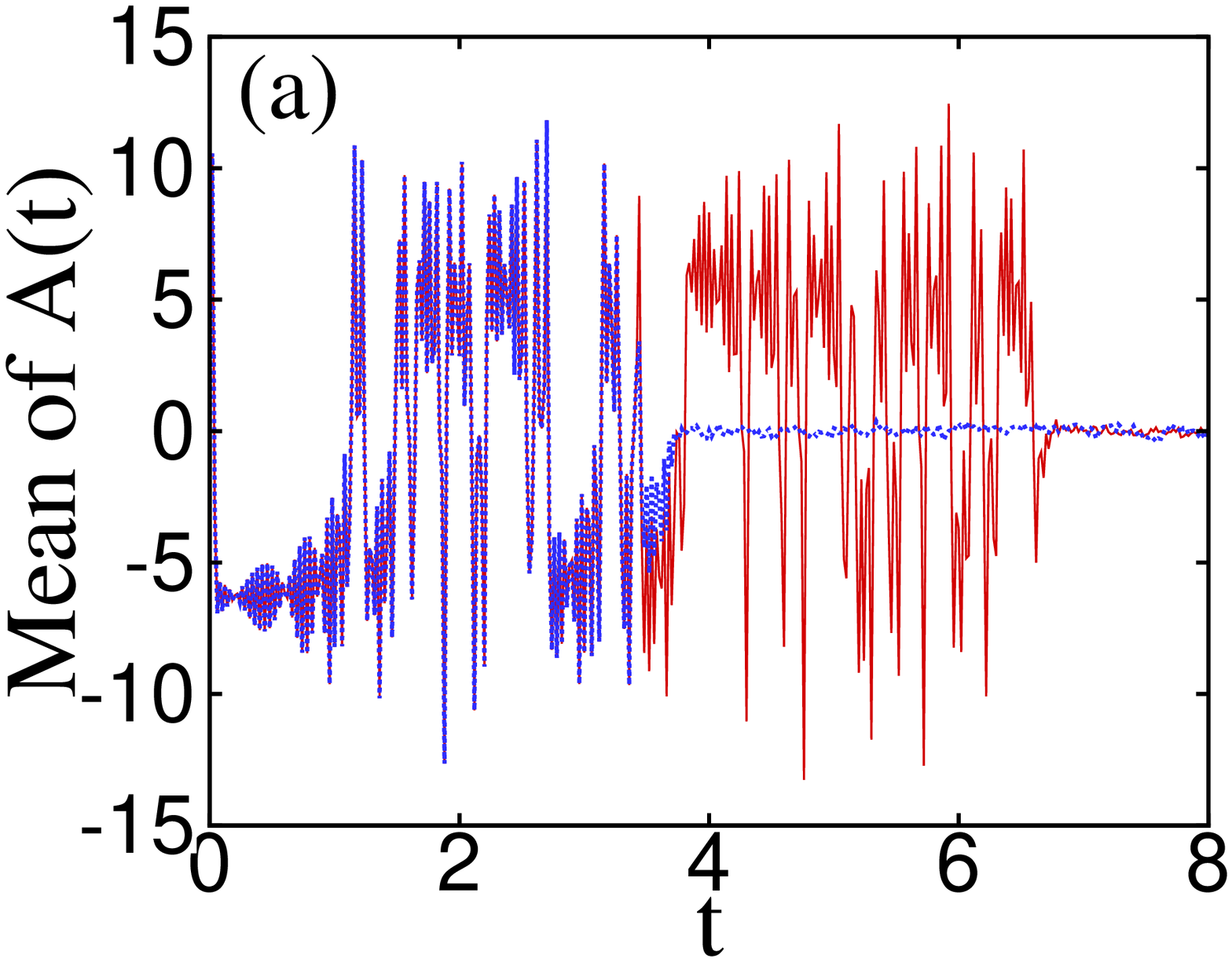}
  \includegraphics[scale=0.3]{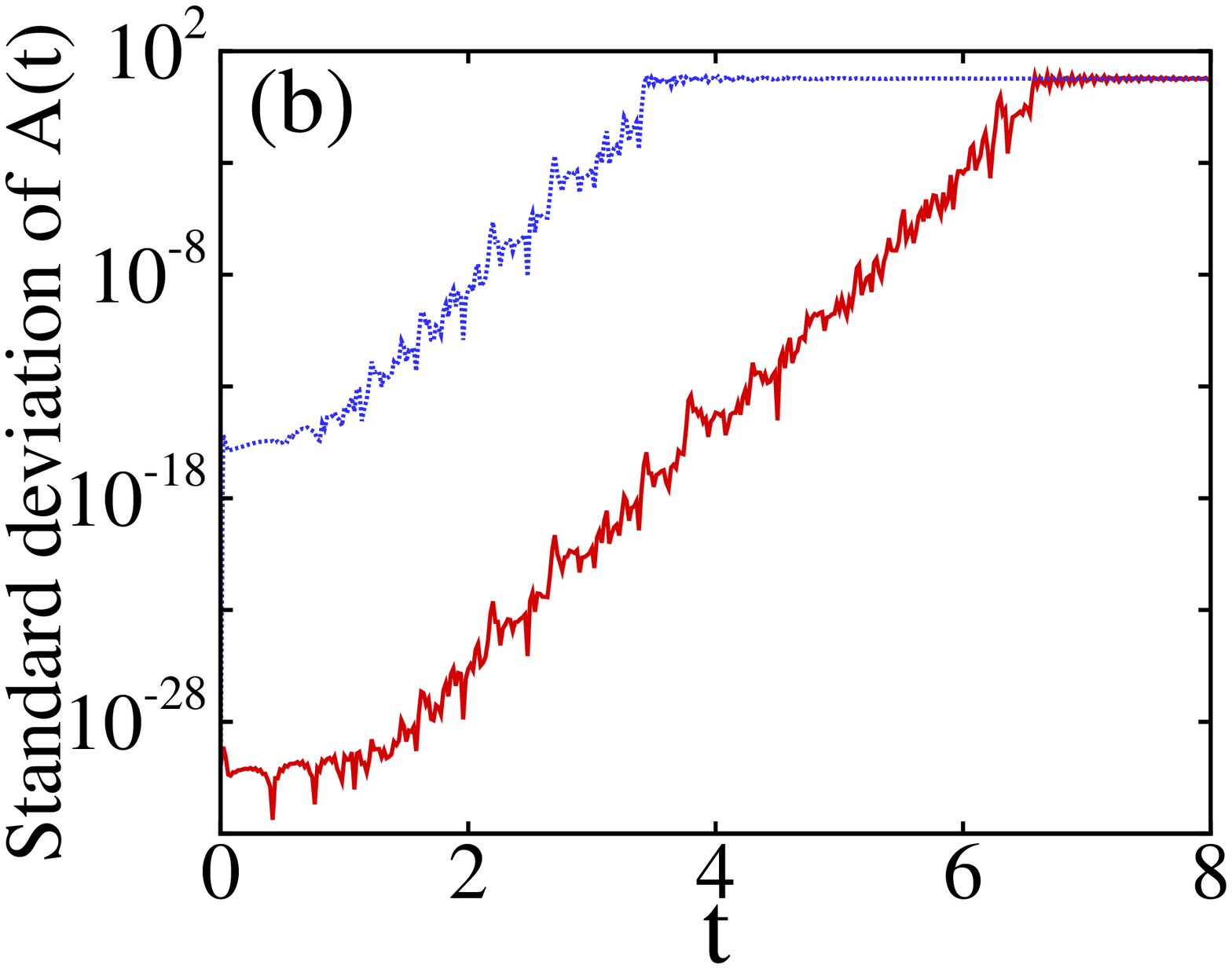}
  \caption{The influence of the round-off error on (a) the mean and (b) the standard deviation of $A(t)$ of the three ODEs' chaotic system (see Eq.(\ref{3eqns}))  using the time-step $h= 10^{-3}$ and the 80th-order Taylor expansion ($M=80$) with the negligible truncation error.   Solid line in red: the reliable CNS results given by the 90 decimal-digit multiple precision for every data; dashed line in blue:  the unreliable results given by means of double precision.}
\label{fig:3eqs-M80K300-roundoff}
\end{figure}

The reliable mean  (the red line, given by $M=80$)  of $A(t)$ of the three ODEs' chaotic dynamic system (Eq.(\ref{3eqns})) is shown in Fig.~\ref{fig:3eqns-M80K300-M10K300}, compared with the unreliable result (the blue line, given by $M=10$).    Note that the reliable mean of $A(t)$ given by the CNS becomes stable when $t>7$, but is time-dependent when $t\leq 7$.   However,  as shown in Fig.~\ref{fig:3eqns-M80K300-M10K300}, the unreliable mean of $A(t)$  given by $M=10$ (with the considerable truncation error) has a noticeable difference from the reliable mean given by the CNS  within  $2.5<t<7$.
It suggests that the truncation error has a great influence on the {\em unsteady} statistical quantities when the chaotic dynamic system is in a transition stage from an equilibrium state to another one.  Note that the mean of $A(t)$ given by $M=10$ agrees well with the reliable CNS result when $t>7$.  Thus,  the truncation error seems to have no influence on the {\em time-independent} statistics of chaotic dynamic systems in an equilibrium state.   The scatter diagrams of the probability distribution function (PDF) in the $A-D$ plane of the three ODEs' chaotic dynamic system (Eq.(\ref{3eqns}))  given by the two numerical schemes (i.e. $M=10$ and $M=80$) are as shown in Fig.~\ref{fig:3eqns-GaussianKDE}, where the PDF is obtained by using the Gaussian kernel density estimator.  Note that  the PDFs given by the two numerical approaches are  almost the same at $t=8.00$.  However, at  $t=6.60$ and $t=6.80$,  the PDFs given by $M=10$ are quite different from those given by the CNS using $M=80$.  This supports our previous conclusion that the truncation error has a significant influence on the unsteady statistics of the chaotic dynamic system, but no influence on steady ones.   This might be the reason why many DNS results for fully developed turbulence agree well with experimental data.

\begin{figure}
  \centering
  \includegraphics[scale=0.8]{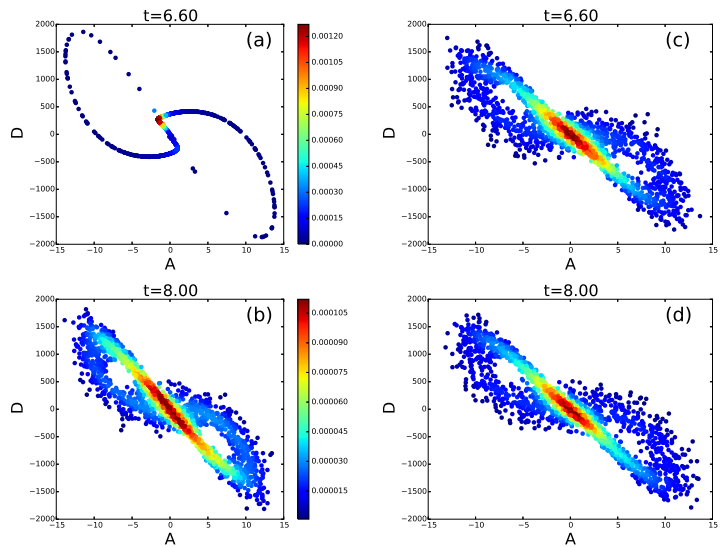}
  \caption{The scatter diagrams of the probability distribution function (PDF) in the $A-D$ plane of the three ODEs' chaotic system (see Eq.(\ref{3eqns}))  at different time.  (a) and (b): reliable results given by the CNS using the 90 decimal-digit multiple precision for every data;  (c) and (d): the unreliable results given by double precision. (color online)}
\label{fig:3eqns-scatter-round-off}
\end{figure}

How about the influence of round-off error on chaotic systems?  It should be emphasized that  double precision is widely used in numerical simulations,  which brings  round-off errors at {\em every} time-step that increase exponentially for chaotic dynamic systems.   In order to investigate the influence of round-off error,  we add a random data  at each time-step with zero mean and the standard deviation  $10^{-16}$, while the 80th-order Taylor expansion ($M=80$) is still used and all data are  expressed in 90 decimal-digits so as to guarantee the negligible truncation error in the considered interval of time $t\in[0,10]$.

Figure~\ref{fig:3eqs-M80K300-roundoff} shows the comparison between the reliable statistics  given by the CNS (using the 90 decimal-digit multiple precision for every data) and the unreliable statistics  given by double precision.   Note that the round-off error has a great influence on the standard deviation of $A(t)$ from the very beginning.    Within $0\leq t\leq 3.5$, the difference of  the standard deviation of $A(t)$  is so small that no obvious separation of the mean is  observed.  However,    at about $t \approx 3.5$ when the round-off error enlarges exponentially so that the standard deviation reach the level of the ``true''  chaotic  solution,  the mean of $A(t)$ becomes  unreliable.    Note that the mean of $A(t)$ given by the double precision becomes {\em time-independent} more early,  at $t \approx 3.5$, which is however wrong in physics, since  the correct mean $A(t)$ of the chaotic system  should become time-independent when $t>7$,  according to our reliable CNS result.    As the system  truly  becomes  unsteady  when $t>7$,  the  round-off  error  has  no  influence on statistics.    Furthermore,  Figure~\ref{fig:3eqns-scatter-round-off} shows the scatter diagram of the probability distribution function (PDF)  in the $A-D$ plane of the three ODEs' chaotic dynamic system (Eq.(\ref{3eqns}))  given by the two different numerical schemes.    Note that  the PDFs  are almost the same at $t=8.00$.  However, the two PDFs are  obviously different at $t=6.60$.   All of these suggest that the round-off error might  have a great influence on the {\em unsteady}  statistics when the chaotic dynamic system is in a transition stage from an equilibrium state to a new one, but has a little impact on steady ones.

\begin{figure}
  \centering
  \includegraphics[scale=0.3]{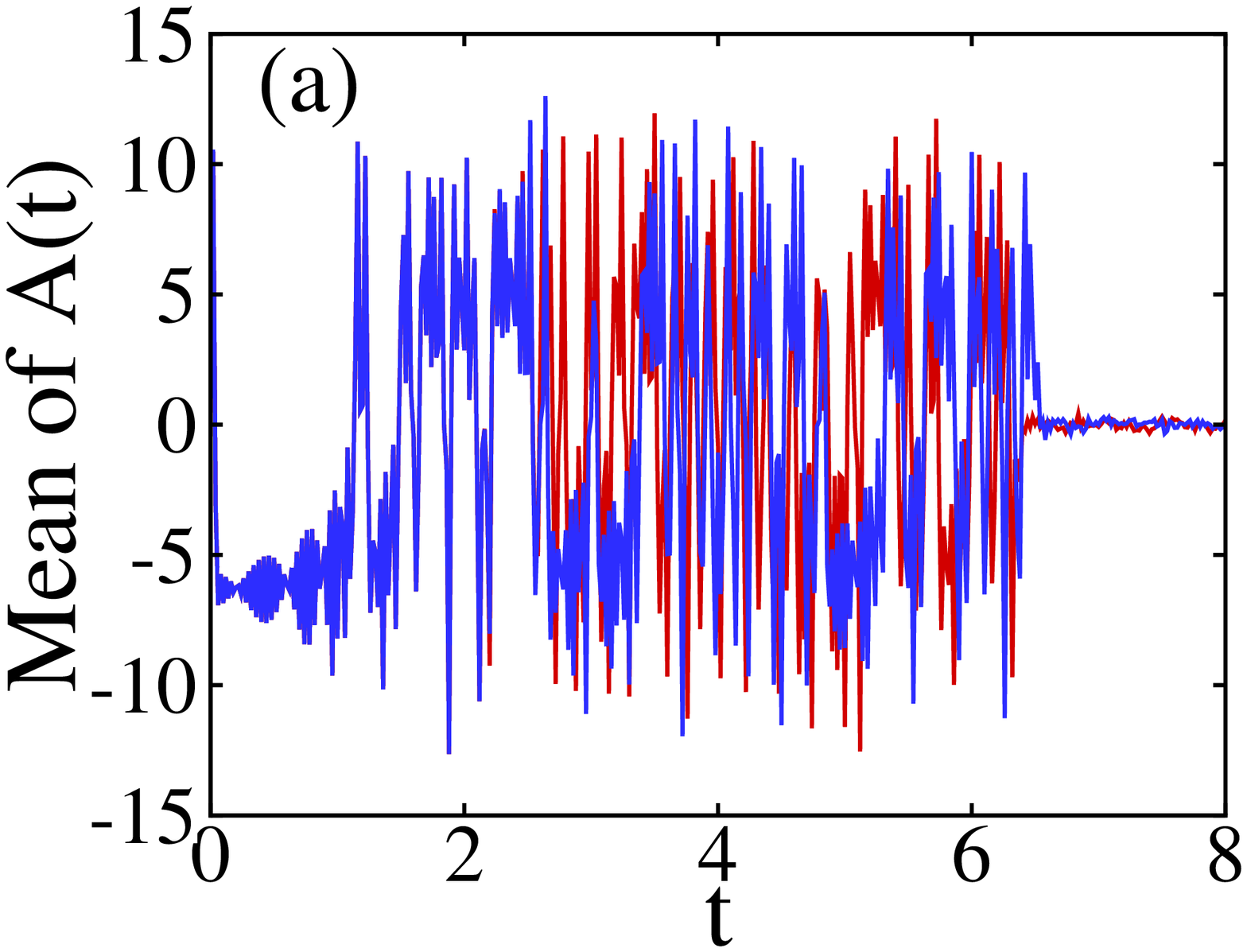}
  \includegraphics[scale=0.3]{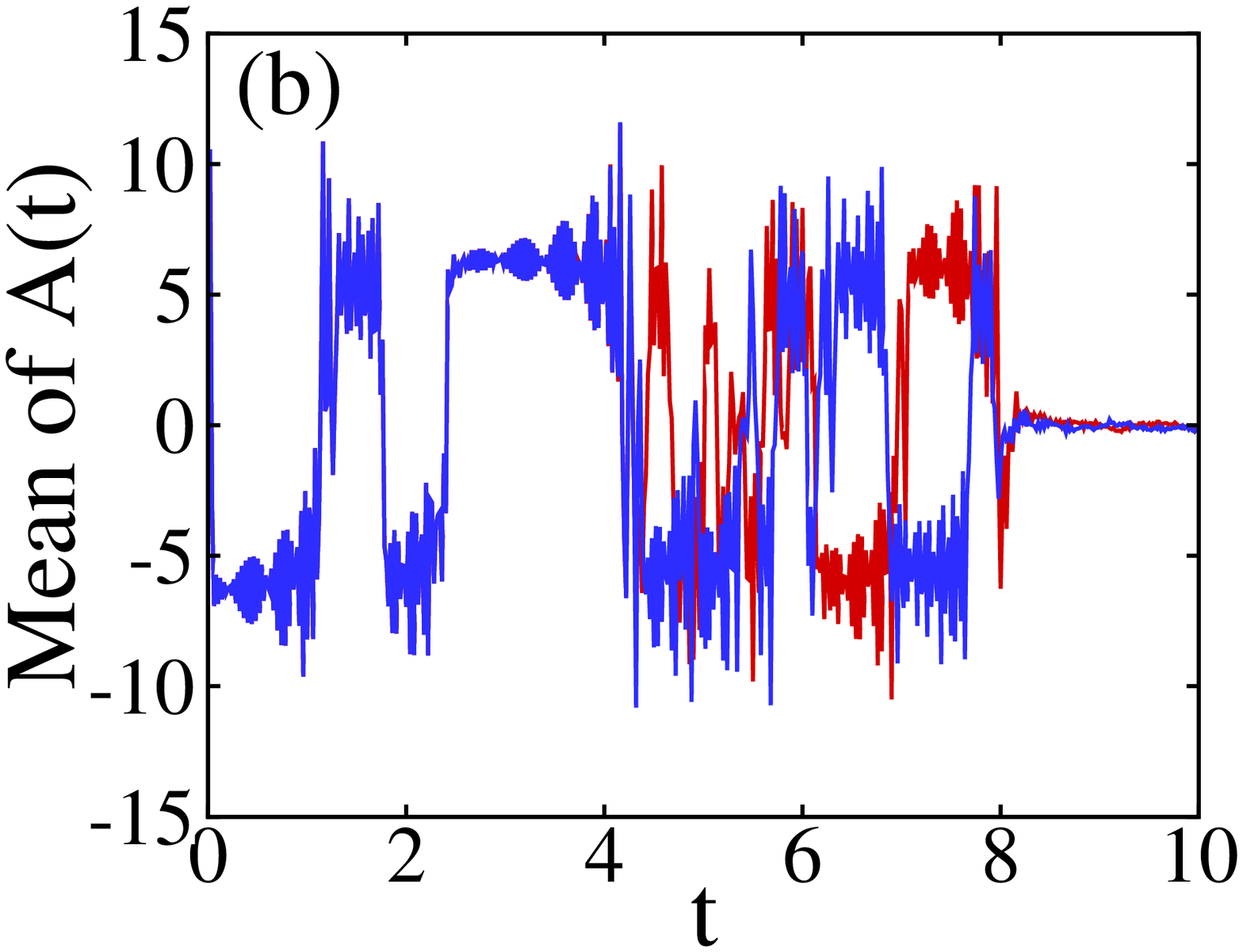}
  \caption{The influence of the truncation error on the mean of $A(t)$ of (a) the five ODEs'  and (b) the seven ODEs' chaotic systems obtained using the time-step $h =10^{-3}$  and the 90 decimal-digit multiple precision for every data with the negligible round-off error.   Solid line in red:  the reliable  results given by the CNS using the 80th-order Taylor series expansion for $t$ (i.e. $M=80$, with the negligible truncation error);  dashed line in blue:  the unreliable result given by the 10th-order Taylor series expansion (i.e. $M=10$, with the considerable truncation error). }
\label{fig:5eqns-7eqns}
\end{figure}

Similarly, we also investigate the influence of numerical noises on computations of statistics of the five ODEs' and seven ODEs'  chaotic dynamic systems,  governed by Eqs.(\ref{5eqns}) and (\ref{7eqns}), respectively,  and obtain the same conclusions qualitatively,  as shown in Fig.~\ref{fig:5eqns-7eqns}.    It suggests that the above-conclusions should have general meanings and would be qualitatively the same even for chaotic systems with large numbers of DOF, say, numerical noises might have a great influence even on  statistic properties  if  chaotic dynamic systems are {\em time-dependent} in statistics.   The same conclusion was also reported for chaotic three-body system\upcite{38} and therefore has general meanings.

\section{Concluding remarks and discussions}

In this paper, we illustrate that convergent (reliable) simulations of chaotic dynamic systems can {\em not} be obtained by means of the traditional numerical algorithms in double precision, even including symplectic integrators, although  they  can  conserve the total energy of the chaotic systems  quite well. So,  conservation of total energy itself  can {\em not} guarantee the reliability of trajectories, Fourier power spectra and autocorrelation functions of chaotic dynamic systems. Such kind of numerical phenomenon lead to some intense arguments about reliability of numerical simulations of chaos\upcite{5,6,8,9}.

These arguments can be calmed down by means of the Clean Numerical Simulation (CNS) that can overcome these defects.   Based on arbitrary order of Taylor series  expansion and data in arbitrary accuracy of multiple precision plus a convergence (reliability) check by means of comparison with an additional simulation with even smaller numerical noises, the CNS provides us a new strategy to reduce numerical noises to any a required level so that numerical noises are negligible in a long (but finite) interval of time.    Using the chaotic  H\'{e}non-Heiles system and the famous three-body problems as example, as illustrate that the CNS can give convergent, reliable chaotic trajectories in a quite long interval of time.   It should be emphasized that, for the H\'{e}non-Heiles system and the three-body problem,  it is  very  important  to  give  an  accurate   prediction of orbits.  Therefore,  the CNS  can indeed  bring us something completely new/different,  although it is a kind of remixing of some known methods/technologies.

The SDIC of chaotic dynamic systems is well-known, which can be traced back to Poincar\'{e}\upcite{1} in 1880s and Lorenz\upcite{2} in 1960s who gave the SDIC a more popular name, i.e. the butterfly-effect.   Even so,  it is nowadays widely believed that statistic properties of chaotic dynamic systems could be correctly obtained, even if convergent chaotic trajectories are impossible by means of traditional algorithms in double precisions.   However, strictly speaking, this is only a kind of {\em conjecture}, say,  an ideal wish.   In this paper, we illustrate that   even  {\em statistic} properties of chaotic systems   can {\em not}  be correctly obtained by means of  traditional numerical algorithms in double precision, as long as these statistics are time-dependent!   The same conclusion was also reported for chaotic three-body system\upcite{38} and therefore has general meanings.

Note that above-mentioned conclusions are based on some simple chaotic systems such as Lorenz equation, the three-body problem, the  Hamiltonian H\'{e}non-Heiles system, and so on.   How about more complicated dynamic systems with an infinite number of dimensions, such as turbulent flows?  Note that the famous Lorenz equation with three degrees of freedom is a simplified model, which can be derived from the exact Navier-Stokes equation describing the two-dimensional Rayleigh-B{\'e}nard convection flow\upcite{61}.  Currently,  Lin {\em et al.} successfully applied the CNS to the two-dimensional Rayleigh-B{\'e}nard convection flow\upcite{39}:  solving  a  dynamic  system  of  16129 (=$127\times 127$)  degrees of freedom  by means of the CNS,  they provided a theoretical evidence that the exact Navier-Stokes equation  has the SDIC and thus is chaotic.    So, it is quite possible that numerical noises might have a great influence on statistic properties of turbulent flows, when these statistics are {\em time-dependent}.   Thus, our CNS results strongly suggest that we had better to be careful on results of statistically unsteady turbulent flows given by the DNS in double precision, although DNS results often agree well with experimental data when turbulent flows are in a statistical stationary state, as reported in\upcite{12,16}.

Obviously,   with negligible numerical noises in a long enough  interval of time,  the CNS could provide us a better and more reliable way to investigate chaotic dynamic systems.  It should be emphasized that, by means of the CNS, we gave, for the first time,  a convergent (reliable) chaotic trajectory of Lorenz equation in a quite long interval [0,10000], which has been never obtained by any traditional algorithms in double precision\upcite{35}.   Besides, the CNS provides, for the first time,  a theoretical evidence\upcite{39} that the Navier-Stokes equations have the SDIC and is chaotic so that micro-level thermal fluctuation becomes the origin of the macroscopic randomness of the Rayleigh-B{\'e}nard turbulent flows, say,  randomness of turbulent flows is self-excited, i.e. out of nothing.    In addition,  hundreds of new periodic orbits of the three-body system with equal mass and zero momentum and over a thousand periodic orbits of planar
three-body system with unequal mass have been found by means of the CNS, for the first time\upcite{40,42}.    Indeed, the CNS can bring us something completely new/different, although it is only a remixing of some known methods/technologies that can be traced back even to Newton.

\section*{Acknowledgment}
We thank the  anonymous  reviewers  for  their  valuable  comments and  suggestions.






\begin{thebibliography}{00}
\bibitem{1}POINCAR\'{E}, J. H.  Sur le probl\`{e}me des trois corps et les \'{e}quations de la dynamique. Divergence des s\'{e}ries de M. Lindstedt. \textit{Acta Mathematica}, \textbf{13}, 1--270 (1890)

\bibitem{2}LORENZ, E. N. Deterministic nonperiodic flow. \textit{Journal of the Atmospheric Sciences}, \textbf{20}(2), 130--141 (1963)

\bibitem{3}LORENZ, E. N. Computational chaos - a prelude to computational instability. \textit{Physica D}, \textbf{15}, 299--317 (1989)

\bibitem{4}LORENZ, E. N.  Computational periodicity as observed in a simple system. \textit{Tellus A}, \textbf{58}, 549--559 (2006)

\bibitem{5}LI, J., ZENG, Q. and CHOU, J. Computational uncertainty principle in nonlinear ordinary differential equations (II): theoretical analysis.  \textit{Science in China (Series E)}, \textbf{44}, 55--74 (2001)

\bibitem{6}TEIXEIRA, J., REYNOLDS, C. and JUDD, K.  Time step sensitivity of nonlinear atmospheric models: Numerical convergence, truncation error growth, and ensemble design. \textit{Journal of the Atmospheric Sciences}, \textbf{64}, 175--188 (2007)

\bibitem{7}QIN S. and LIAO S. Influence of round-off errors on the reliability of numerical simulations of chaotic dynamic systems. \textit{Journal of Applied Nonlinear Dynamics}(accepted) (Preprint arXiv:1707.04720)

\bibitem{8}YAO, L. and HUGHES, D.  Comment on ``computational periodicity as observed in a simple system" by Edward N. Lorenz (2006). \textit{Tellus A}, \textbf{60}, 803--805 (2008)

\bibitem{9}LORENZ, E. N.  Reply to comment by L.-S. Yao and D. Hughes. \textit{Tellus A}, \textbf{60}, 806--807 (2008)

\bibitem{10}ALBERS, T. and RADONS, G. Weak ergodicity breaking and aging of chaotic transport in Hamiltonian systems. \textit{Physical Review Letters}, \textbf{113}(18), 184101 (2014)

\bibitem{11}HUYNH, H. N., NGUYEN,  T. P. T. and CHEW, L. Y. Numerical simulation and geometrical analysis on the onset of chaos in a system of two coupled pendulums. \textit{Communications in Nonlinear Science and Numerical Simulation}, \textbf{18}(2),291--307 (2013)

\bibitem{12}LEE, M. and MOSER, R. D.  Direct numerical simulation of turbulent channel flow up to to ${R}e_{\tau}\approx5200$. \textit{Journal of Fluid Mechanics}, \textbf{774}, 395--415 (2015)

\bibitem{13}WANG, J.  LI, Q.  and E, W. N., Study of the instability of the Poiseuille flow using a thermodynamic formalism. \textit{Proceedings of the National Academy of Sciences}, \textbf{112}(31), 9518--9523 (2015)

\bibitem{14} AVILA, K., MOXEY, D., de LOZAR, A., AVILA, M., BARKLEY, D., and HOF, B. The onset of turbulence in pipe flow. \textit{Science}, \textbf{333}(6039), 192--196 (2011)

\bibitem{15}DEIKE, L., FUSTER, D., BERHANU, M., and FALCON, E. Direct numerical simulations of capillary wave turbulence. \textit{Physical Review Letters}, \textbf{112}(23), 234501 (2014)

\bibitem{16}KIM, J., MOIN, P., and MOSER, R. Turbulence statistics in fully developed channel flow at low Reynolds number. \textit{Journal of Fluid Mechanics}, \textbf{177}, 133--166 (1987)

\bibitem{17}YEE, H., TORCZYNSKI, J., MORTON, S., VISBAL, M. and  SWEBY, P. On spurious behavior of CFD simulations. \textit{International Journal for Numerical Methods in Fluids}, \textbf{30}(6), 675--711 (1999)

\bibitem{18} WANG, L. P., and ROSA, B. A spurious evolution of turbulence originated from round-off error in pseudo-spectral simulation. \textit{Computers \& Fluids}, \textbf{38}(10), 1943--1949 (2009).

\bibitem{19}YEE, H. C., SWEBY, P. K., and GRIFFITHS, D. F.  Dynamical approach study of spurious steady-state numerical solutions of nonlinear differential equations. I. The dynamics of time discretization and its implications for algorithm development in computational fluid dynamics. \textit{Journal of Computational Physics}, \textbf{97}(2), 249--310 (1991)

\bibitem{20}YEE, H. C., and SWEBY, P. K.  Dynamical approach study of spurious steady-state numerical solutions of nonlinear differential equations II. Global asymptotic behavior of time discretizations. \textit{International Journal of Computational Fluid Dynamics}, \textbf{4}(3-4), 219--283 (1995)

\bibitem{21}KRYS'KO, V.A., AWREJCEWICZ, J. and BRUK, V.M. On the solution of a coupled thermo-mechanical problem for non-homogeneous Timoshenko-type shells. \textit{Journal of Mathematical Analysis and Applications}, \textbf{273}(2), 409--416 (2002)

\bibitem{22}AWREJCEWICZ, J. and KRYSKO, V.A. Nonlinear coupled problems in dynamics of shells. \textit{International Journal of Engineering Science}, \textbf{41}(6), 587--607 (2003)

\bibitem{23}AWREJCEWICZ, J., KRYSKO, V.A. and KRYSKO, A.V. Complex parametric vibrations of flexible rectangular plates. \textit{Meccanica}, \textbf{39}(3), 221--244 (2004)

\bibitem{24}AWREJCEWICZ, J., KRYSKO, A.V., ZHIGALOV, M.V., SALTYKOVA, O.A. and KRYSKO, V.A. Chaotic vibrations in flexible multi-layered Bernoulli-Euler and Timoshenko type beams. \textit{Latin American Journal of Solids and Structures}, \textbf{5}(4), 319--363 (2008)

\bibitem{25}AWREJCEWICZ, J., KRYSKO, A.V., KUTEPOV, I.E., ZAGNIBORODA, N.A., DOBRIYAN, V. and KRYSKO, V.A. Chaotic dynamics of flexible Euler-Bernoulli beams. \textit{Chaos}, \textbf{34}(4), 043130 (2014)

\bibitem{26}KRYSKO, A.V., AWREJCEWICZ, J., SALTYKOVA, O.A., ZHIGALOV, M.V. and KRYSKO, V.A. Investigations of chaotic dynamics of multi-layer beams using taking into account rotational inertial effects. \textit{Communications in Nonlinear Science and Numerical Simulation}, \textbf{19}(8), 2568--25890 (2014)

\bibitem{27}AWREJCEWICZ, J., KRYSKO JR, V. A., PAPKOVA, I. V., KRYLOV, E. Y., and KRYSKO, A. V.  Spatio-temporal non-linear dynamics and chaos in plates and shells. \textit{Nonlinear Studies}, \textbf{21}(2), 313--327 (2004)

\bibitem{28}AWREJCEWICZ, J., KRYSKO, A. V., ZAGNIBORODA, N. A., DOBRIYAN, V. V., and KRYSKO, V. A. On the general theory of chaotic dynamic of flexible curvilinear Euler-Bernoulli beams. \textit{Nonlinear Dynamics}, \textbf{85}(4), 2729--2748 (2016)

\bibitem{29}AWREJCEWICZ, J., KRYSKO, A.V., PAPKOVA, I.V., ZAKHAROV, V.M., EROFEEV, N.P., KRYLOVA, E.Y., MROZOWSKI, J. and KRYSKO, V.A. Chaotic dynamics of flexible beams driven by external white noise . \textit{Mechanical Systems and Signal Processing}, \textbf{79}, 225--253 (2016)

\bibitem{30}AWREJCEWICZ, J., KRYSKO, A.V., EROFEEV, N.P., DOBRIYAN, V., BARULINA, M.A. and KRYSKO, V.A. Quantifying chaos by various computational methods. Part 1: Simple systems. \textit{Entropy}, \textbf{20}(3), 175 (2018)

\bibitem{31}AWREJCEWICZ, J., KRYSKO, A.V., EROFEEV, N.P., DOBRIYAN, V., BARULINA, M.A. and KRYSKO, V.A. Quantifying chaos by various computational methods. Part 2: Vibrations of the Bernoulli-Euler beam subjected to periodic and colored noise. \textit{Entropy}, \textbf{20}(3), 170 (2018)

\bibitem{32}LIAO, S.  On the reliability of computed chaotic solutions of non--linear differential equations. \textit{Tellus A}, \textbf{61}(4), 550--564 (2009)

\bibitem{33}WANG, P., LI, J., and LI, Q.  Computational uncertainty and the application of a high-performance multiple precision scheme to obtaining the correct reference solution of Lorenz equations. \textit{Numerical Algorithms}, \textbf{59}(1), 147--159 (2012)

\bibitem{34}LIAO, S.  Physical limit of prediction for chaotic motion of three-body problem. \textit{Communications in Nonlinear Science and Numerical Simulation}, \textbf{19}(3), 601--616 (2014)

\bibitem{35}LIAO, S., and WANG, P.  On the mathematically reliable long-term simulation of chaotic solutions of Lorenz equation in the interval [0, 10000]. \textit{Science China Physics, Mechanics and Astronomy}, \textbf{57}(2), 330--335 (2014)

\bibitem{36}LIAO, S.  Can we obtain a reliable convergent chaotic solution in any given finite interval of time? \textit{International Journal of Bifurcation and Chaos}, \textbf{24}(09), 1450119 (2014)

\bibitem{37}LI, X., and LIAO, S.  On the stability of the three classes of Newtonian three--body planar periodic orbits. \textit{Science China Physics, Mechanics \& Astronomy}, \textbf{57}(11), 2121--2126 (2014)

\bibitem{38}LIAO, S., and LI, X.  On the Inherent Self-Excited Macroscopic Randomness of Chaotic Three-Body Systems. \textit{International Journal of Bifurcation and Chaos}, \textbf{25}(09), 1530023 (2015)

\bibitem{39}LIN, Z., WANG, L., and LIAO, S. On the origin of intrinsic randomness of Rayleigh-B{\'e}nard turbulence. \textit{Science China Physics, Mechanics \& Astronomy}, \textbf{60}(1), 014712(2017)

\bibitem{40}LI, X. and LIAO, S.  More than six hundred new families of Newtonian periodic planar collisionless three-body orbits. \textit{Science China Physics, Mechanics \& Astronomy}, \textbf{60}(12), 129511 (2017)

\bibitem{41}LIAO, S.  On the clean numerical simulation (cns) of chaotic dynamic systems. \textit{Journal of Hydrodynamics}, \textbf{29}(5), 729--747 (2017)

\bibitem{42}LI, X., JING, Y. and LIAO, S.  Over a thousand new periodic orbits of planar three-body system with unequal mass. \textit{Publications of the Astronomical Society of Japan(accepted)},(2018)

\bibitem{43}BARTON, D., WILLERS, I. M., and ZAHAR, R. V. M.  The automatic solution of systems of ordinary differential equations by the method of Taylor series. \textit{The Computer Journal}, \textbf{14}(3), 243--248 (1971)

\bibitem{44}CORLISS, G., and LOWERY, D.  Choosing a stepsize for Taylor series methods for solving ODE'S. \textit{Journal of Computational and Applied Mathematics}, \textbf{3}(4), 251--256 (1977)

\bibitem{45}CORLISS, G., and CHANG, Y. F. Solving ordinary differential equations using Taylor series. \textit{ACM Transactions on Mathematical Software}, \textbf{8}(2), 114--144 (1982)

\bibitem{46}JORBA, {\`A}., and ZOU, M.  A software package for the numerical integration of ODEs by means of high-order Taylor methods. \textit{Experimental Mathematics}, \textbf{14}(1), 99--117 (2005)

\bibitem{47}BARRIO, R., BLESA, F., and LARA, M.  VSVO formulation of the Taylor method for the numerical solution of ODEs. \textit{Computers \& Mathematics with Applications}, \textbf{50}(1-2), 93--111 (2005)

\bibitem{48}PORTILHO, O.  MP--A multiple precision package. \textit{Computer Physics Communications}, \textbf{59}, 345--358 (1990)

\bibitem{49}SUN, B. Kepler¡¯s third law of n-body periodic orbits in a Newtonian gravitation field. \textit{Science China Physics, Mechanics \& Astronomy}, \textbf{61}, 054721 (2018)

\bibitem{50}FRISCH, A., MARK, M., AIKAWA, K., FERLAINO, F., BOHN, J. L., MAKRIDES, C., PETROV, A., and KOTOCHIGOVA, S.  Quantum chaos in ultracold collisions of gas-phase erbium atoms. \textit{Nature}, \textbf{507}(7493), 475 (2014)

\bibitem{51}SUSSMAN, G. J., and WISDOM, J.  Chaotic evolution of the solar system. \textit{Science}, \textbf{257}(5066), 56--62 (1992).

\bibitem{52}MCLACHLAN, R. I., MODIN, K., and VERDIER, O.  Symplectic integrators for spin systems. \textit{Physical Review E}, \textbf{89}(6), 061301 (2014)

\bibitem{53}LASKAR, J., and ROBUTEL, P.  High order symplectic integrators for perturbed Hamiltonian systems. \textit{Celestial Mechanics and Dynamical Astronomy}, \textbf{80}(1), 39--62 (2001)

\bibitem{54}QIN, H., and GUAN, X.  Variational symplectic integrator for long-time simulations of the guiding-center motion of charged particles in general magnetic fields. \textit{Physical Review Letters}, \textbf{100}(3), 035006 (2008)

\bibitem{55}FARR{\'E}S, A., LASKAR, J., BLANES, S., CASAS, F., MAKAZAGA, J., and MURUA, A.  High precision symplectic integrators for the solar system. \textit{Celestial Mechanics and Dynamical Astronomy}, \textbf{116}(2), 141--174 (2013).

\bibitem{56}FOREST, E., and RUTH, R. D.  Fourth-order symplectic integration. \textit{Physica D: Nonlinear Phenomena}, \textbf{43}(1), 105--117 (1990)

\bibitem{57}YOSHIDA, H. Construction of higher order symplectic integrators. \textit{Physics Letters A}, \textbf{150}(5--7), 262--268 (1990)

\bibitem{58}H{\'E}NON, M., and HEILES, C.  The applicability of the third integral of motion: some numerical experiments. \textit{The Astronomical Journal}, \textbf{69}, 73 (1964)

\bibitem{59}SPROTT, J. C. Elegant chaos: algebraically simple chaotic flows. \textit{World Scientific} (2010)

\bibitem{60}LIAO, S. On the numerical simulation of propagation of micro-level inherent uncertainty for chaotic dynamic systems. \textit{Chaos, Solitons \& Fractals}, \textbf{47}, 1--12 (2013)

\bibitem{61}SALTZMAN, B.  Finite amplitude free convection as an initial value problem--I.\textit{Journal of the Atmospheric Sciences}, \textbf{19}(4), 329--341 (1962)
\end{thebibliography}






\end{document}